# Aristotle Cloud Federation: Container Runtimes Technical Report

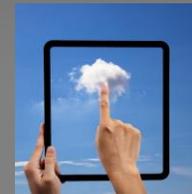

Peter Z. Vaillancourt*  Rich Wolski**
Bennett Wineholt*  Christopher R. Myers*
Tristan J. Shepherd*  Ben Trumbore*
Sara C. Pryor*  Resa Reynolds*
Jeffrey Lantz*  Jodie Sprouse*
Richard Knepper*  David Lifka*

*Cornell University **University of California, Santa Barbara

A National Science Foundation-sponsored container runtimes investigation was conducted by the Aristotle Cloud Federation to better understand the challenges of selecting and using Docker, Singularity, and X-Containers. The main goal of this investigation was to identify the "pain points" experienced by users when selecting and using containers for scientific research and to share lessons learned. Application performance characteristics are included in this report as well as user experiences with Kubernetes and container orchestration on cloud and HPC platforms. Scientists, research computing practitioners, and educators may find value in this report when considering the use and/or deployment of containers or when preparing students to meet the unique challenges of using containers in scientific research.

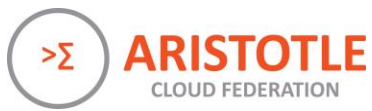

NSF Grant No. OAC-1541215
September 24, 2021

# Contents

# 1.0 Introduction

The NSF-funded Aristotle Cloud Federation project [1] is a Data Infrastructure Building Blocks program project intended to allow institutions the ability to share on-premise computing resources and improve the science usage of cloud technologies in general. Cloud systems were deployed at Cornell University, the University at Buffalo, and the University of California, Santa Barbara, and allocations, accounting, and cloud metrics were implemented. Strategic science use cases are supported with a rich set of open-source software and cloud usage modalities, including Virtual Machine snapshots of complex software systems, dynamically-sized and scheduled compute clusters, optimized frameworks for query-based exploration, modeling, and analysis, and application containers.

Application containerization and distribution technologies developed rapidly during the course of the Aristotle project. Today, application containers provide a level of segregation from the underlying operating system that allows for specific libraries, application versions, and data separate from the host system to remain in the container independent of the host system. The Docker container runtime [2], for example, is popular in the private sector because of its ability to rapidly implement and deploy codes. The research community has been investigating and, in some cases, implementing container technologies for similar reasons. Faster development cycles and portability from laptop to clouds, and even supercomputers, are very appealing despite the challenges of implementing containers.

Based on commentary from project reviewers and the desire for portability on the part of Aristotle use case scientists (astronomers searching for pulsars and Fast Radio Bursts [3], atmospheric scientists optimizing wind turbine production [4], etc.), the Aristotle team took on a supplemental activity to review and compare application container implementations and runtimes. Our experiences and lessons learned are shared in this report for scientists and cyberinfrastructure (CI) support professionals who are considering adding application containers to their research portfolios.

The "Aristotle Cloud Federation: Container Runtimes Technical  Report" provides an overview of three container runtimes—Docker, Singularity [5], and X-Containers [6]—and shares our experiences implementing them on multiple platforms. Implementation of containers on a project that required Weather Research and Forecasting (WRF) Modeling software [7] was a particularly good use case challenge because it required real-world constraints such as needing to plan for analyses of large data sets, separating complex build environments from execution environments, and instrumenting containers to understand performance characteristics.

The report opens with an overview of the container runtimes selected for comparison and the platforms selected to run the application containers on—Aristotle Red Cloud, Stampede2, Amazon Web Services, Google Cloud Platform, and Microsoft Azure. Our container implementation experiences, Kubernetes orchestration, and HPL as a test application are described next. We then compare each of the container runtimes and Kubernetes for ease of use. Finally, our rationale for selecting WRF as a use case is shared, followed by our implementation experiences and the challenge of orchestrating WRF. We conclude by providing the results of a few of our performance experiments.

**Citation**

Any opinions, findings, and conclusions or recommendations expressed in this material are those of the authors and do not necessarily reflect the views of the National Science Foundation.

Lead author contact: Peter Z. Vaillancourt at pzv2@cornell.edu.

**Acknowledgements**

We wish to thank the project participants who provided computational resources or gave their time to complete this investigation, including Amazon Web Services, Google, Microsoft, the Texas Advanced Computing Center, and Exotanium co-founders Hakim Weatherspoon and Zhiming Shen. This material is based upon work supported by the National Science Foundation under Grant No. 1541215 (https://federatedcloud.org/).

Edited by: P. Redfern

## 2.0 Container Technology Comparison

In this section, we aim to provide the larger picture of the technological and implementation details for the explored container runtimes and orchestration, as well as an analysis of some of the features from an ease of use perspective. We detail the practical considerations that users and developers of scientific software applications face when considering using containers, as well as some details on how to get started with implementing container software on multiple platforms. We begin with a general overview of the technologies used (containers, platforms, and orchestration), discuss implementation considerations for scientific applications with each container type including optional orchestration, and summarize the main takeaways from this comparison. A focus on HPC applications in general pervades our analysis of container runtimes and tools. A detailed analysis of containerizing and orchestrating WRF follows in Section 3.0: "Containerization and Orchestration for WRF."

### 2.1 Technology Overview

#### 2.1.1 Container Runtimes

**Docker**

Docker is an industry standard container runtime which allows users to bundle up software components in a reproducible system environment. Based on Linux cgroups, users commonly choose a base image—often a minimal operating system (OS) offering—for their container and package files and applications to be included in their Docker image using a customizable scripted build process. The Docker build process copies in files, compiles application binaries, and combines the system layers and specified software layers into a container image. Docker images thus built can then be run as a container instance on the same or other host systems provided that a compatible kernel, hypervisor, and container runtime are present.

Unlike full Virtual Machines (VMs), Docker containers do not encapsulate the entire OS, rather they rely on the host kernel to execute through the Docker daemon host process. By default, this daemon requires root access to the host system that it runs on which has been a well-publicized source of concern for the security of the host system. These concerns have led much of the HPC community to avoid adoption of Docker containers. Recently, Docker has begun supporting a "rootless" mode to enhance the security of the runtime environment and systems hosting it [8]. As of this writing, however, HPC systems still largely rely on alternative container runtimes, some of which can run Docker images in a modified fashion.

Docker has been widely adopted by industry to simplify application building and deployment across distributed environments with a broad ecosystem of open source and enterprise tools for building, distributing, running, and managing services. Free for the hosting of public container images, Docker Hub [9] is a valuable service for the developer community which allows Docker users to find examples of similar applications to base their builds on, inspect build script codes, and quickly deploy sample applications to test and even accelerate their own distribution of publicly-shared software applications. Furthermore, there are a wide array of development, CI/CD, and production tools available for use with Docker containers and the Docker runtime environment.

## Singularity

Singularity is an image format and container runtime based on the Docker container concept, but adapted for secure and portable usage on many platforms ranging from laptops to supercomputers. To avoid the security concerns originating with the root user access issue in Docker, Singularity was developed to run containers in userspace. In this way, anyone with access to the Singularity container has only the permissions of the user who ran the container and no root filesystem access. This enhanced security model encouraged more container adoption within the HPC community, especially since Singularity is designed to be interoperable with Docker.

Users can either import Docker images for conversion or build Singularity images through a process similar to Docker. The user can provide software, scripts to run, and files to include while packaging applications into the highly portable Singularity Image Format (SIF) [10]. The Singularity build process expands upon Docker in a number of ways, including integrated build tests and cryptographic signing of images using GPG keys. Once a Singularity image has been built, the SIF can be interacted with much like a regular file or the image can be hosted on a remote registry similar to Docker.

The XSEDE community has adopted the Singularity runtime as the method of containerization on HPC systems. The seamless integration of Singularity with Slurm [11] enables significant computational advantages such as ease of scaling and portability. Unfortunately, the demise of Singularity Hub [12] has removed the only free container hosting option with an unlimited storage limit. The Sylabs Cloud Hosting site [13] allows free hosting for up to 11GBs of container images per user, which is very limiting for users with large codebases, such as those present in some HPC communities. For more information on the similarities and differences between Docker and Singularity container runtimes, see Section 2.2.2 "Singularity" on page 13 or consult the Singularity User Guide [14].

## X-Containers

X-Containers [15] is a new container architecture that supports strong security isolation, high performance, and low cost running of application containers in the cloud. It started as a research project at Cornell University and is now maintained by Exotanium, Inc., a start-up that completed the NSF I-Corps program and, today, offers solutions for cloud resource optimization [16]. X-Containers can be deployed seamlessly in existing container orchestration platforms and automatically optimize the performance of an application without the need to modify source code or recompile.

X-Containers is also one of the underlying virtualization solutions for Exotanium's xSpot. X-Spot is a cloud resource optimization and management platform. It uses artificial intelligence and machine learning to minimize cloud computing costs by automatically migrating application containers between VM instances without terminating the container runtimes. This allows a set of containers to be run without interruption on the fewest, smallest instances while taking advantage of spot market pricing.

We tested X-Containers using a proprietary installation hosted on the Aristotle Red Cloud computing cluster. Our evaluations were only concerned with the user experience and any overhead incurred beyond the costs one would expect when running with standard Docker containers.

**Other Container Runtimes**

There are many other container runtimes in active development such that choices must be made to limit the scope of an investigation. In this study, we assessed the container runtimes listed above and their foundational and development tools for their value to scientific software users, especially those utilizing HPC methods. For further evaluation of other runtimes, we direct the curious reader to other container runtime surveys [17] [18] [19] [20] that have either a different user focus or may fail to represent the container runtimes discussed here.

## 2.1.2 Test Environments

**Aristotle Red Cloud**

Red Cloud is a private cloud and Aristotle Cloud Federation resource managed by the Cornell Center for Advanced Computing (CAC) [21] [22] in their data center facility located on the Ithaca, NY campus of Cornell University. It provides VM instance hosting services, block storage, and related management capabilities utilizing OpenStack [23]. Red Cloud features VMs with up to 224GBs memory and 28 virtual central processing unit (vCPU) cores per node instance with no oversubscription to provide better dedicated hardware utilization on the more than 1,400 vCPUs in the cluster [24].

Previously, we developed and tested Docker containers for Aristotle science use cases on Red Cloud VMs [25]. We also deployed an HPC-style self-scaling Virtual Cluster and tested Singularity containers on Red Cloud [26]. This past work was leveraged for the deployment of containerized jobs discussed in this study. Henceforth, we will call this testing environment "Aristotle" for simplicity.

While familiarizing ourselves with a new technology or application, it became common practice to manually provision an Aristotle VM to run code and attempt the installation of tools. The benefit of Aristotle VMs is they provide a controlled cost environment that can be used for development until installation, workflow configuration, and any long running code compilation that can be performed ahead of time.

**Stampede2**

The Stampede2 supercomputer, located at The University of Texas at Austin's Texas Advanced Computing Center (TACC), is number 35 on the June 2021 TOP500 list [27]. Stampede2 consists of a mix of 1,736 Intel Xeon Platinum 8160 nodes ("Skylake" or "SKX") and 4,200 Intel Xeon Phi 7250 nodes ("Knights Landing" or "KNL"). The SKX nodes were used for numerical experiments as part of this project, with hardware support from TACC via an allocation of time on Stampede2 and systems support through the TACC ticket system. Each SKX node on Stampede2 consists of 48 cores (with 2 hardware threads per core, for a total of 96 hardware threads per node) and 192GB DDR4 RAM. The interconnect is a 100Gb/sec Intel Omni-Path network with a 28/20 oversubscription, supporting fast communication for Message Passing Interface (MPI) jobs running on multiple nodes. Stampede2 mounts three shared Lustre file systems that are accessible from all the nodes. Job scheduling on this large shared resource is coordinated using Slurm, which also provides utilities to query job statistics after they have been run. Stampede2 has been a useful resource for this project since it enabled us to run the same applications both on bare metal and in Singularity containers, submitted in both cases through the Slurm scheduler.

Access to software on Stampede2 is organized primarily using the Lmod system to enable users to configure their environments for particular tasks [28]. The TACC system pre-loads a number of useful modules into the user environment by default to ease entry into common computational tasks, and many of these modules were leveraged for this work. Additional modules that are loaded reconfigure the environment to run different types or versions of packages. Of particular importance for this work is the TACC-maintained wrapper on the Singularity container runtime for running container jobs on Stampede2, tacc-singularity/3.7.2. This module must be loaded before any Singularity commands can be used on the system, and the TACC configuration of the Singularity runtime installation affects which features are available to the user.

Loading the modules that provide access to the Intel compilers and Intel MPI libraries was also necessary. This was due to WRF use case requirements and the need for a performant MPI library, which was limited to Intel-supported options that work with the Intel Omni-Path interconnect. Matching versions with other dependencies of our use case determined that we would use version 18 of the Intel compilers and Intel MPI library, though MVAPICH2 was another possible option. Further details regarding dependencies and job configuration on Stampede2 are described in Section 3.2: "WRF Use Case Description" located on page 23.

**Public Cloud**

For this work, we explored deployments of containerized applications in three public clouds: Amazon Web Services (AWS) [29], Google Cloud Platform (GCP) [30], and Microsoft Azure cloud computing services (shortened to "Azure" henceforth) [31].

Each of these public cloud providers offers a wide range of services including the deployment of VMs with various optimized configurations, managed container services, managed cluster services, storage, networking, and much more. For testing within this project, we favored the deployment of cloud-based clusters that could be performed via automation and that were multi-cloud capable to avoid vendor lock-in that might bias our assessment of public cloud container runtime and orchestration capabilities. In some cases, we deployed single instances where it was sufficient and cost-effective to test application containers, but otherwise we focused on the deployment of HPC-style virtual clusters of VMs using Kubernetes [32] orchestration and related tools which are detailed in subsequent sections of this report.

**2.1.3 Kubernetes**

Kubernetes is an open source system for automating the deployment, scaling, and management of containerized applications, and thereby can support the orchestration of container services on virtual clusters deployed in the cloud. Different cloud providers support different variants of the Kubernetes ecosystem. Given our focus on cloud resources in AWS, GCP, and Azure, we made use of the following Kubernetes management systems: Amazon Elastic Kubernetes Service (EKS) [33], Google Kubernetes Engine (GKE) [34], and Azure Kubernetes Service (AKS) [35]

*Why orchestrate?*

In order to perform large scale simulations on multiple VM instances in the style of an HPC cluster in the cloud, orchestration software is necessary. Additionally, orchestration enables the execution of consistent benchmarks to gauge expected performance in a reasonable amount of operator time. The technologies used and amount of scripting and automation that can be usefully leveraged varies considerably based on the science to be performed and cloud platform used. Since we focused our study on HPC applications that are assumed to benefit from cluster computing, the orchestration considerations for other scientific applications may vary from ours.

On some computing resources, deeper system settings are pre-configured which can allow researchers to focus more on the aspects related to running jobs that perform their desired computation. Capable computational support staff can utilize orchestration software to facilitate researcher experiences, but some conversations may be necessary to help train researchers in this technology and set expectations about its capabilities. While training takes time, it is likely necessary considering the plethora of ever-evolving cloud services and tools researchers have to choose from.

*Why Kubernetes orchestration?*

As soon as scaling up a computational task reaches a point that it provides benefits from running containers on more than one node or VM instance, a number of orchestration challenges arise. These challenges include host VMs maintaining awareness of one another's network addresses so long as they are available, coordinating and authenticating access credentials, and updating software container components as new versions emerge. The challenges increase when an application leverages distributed memory programming, such as MPI, as is common in HPC applications. If application software versions or host system configurations drift due to a lack of standardized containers deployed via orchestration or instrumented with automated checks and controls, large scale simulation runs may exhibit the effects of those misconfigurations and suffer abnormal performance or fail to run at the desired scale.

Recognizing these concerns, Google developed Kubernetes container orchestration software [36] and released it as open source. Kubernetes is an industry standard [37] for orchestrating services composed of containers spread across multiple VMs and provides efficient resource utilization, distributed communication, reliable restarts, and other desirable features. Several of Kubernetes features are useful to our work, including managing cloud provider credentials, scaling to larger cluster sizes, and producing distributed logging and performance monitoring data. Kubernetes is a logical choice for multiple node applications that require cloud provider flexibility and the ability to adapt between different cloud implementations.

On the other hand, each public cloud provider supports their own versions of Kubernetes that are tailored to their platform. This means some reconfiguration may be necessary to include cloud provider specifics when migrating from one cloud to another. An additional consideration is data persistence. While data persistence can be achieved via a Network File System (NFS) [38] with a dedicated Kubernetes pod, the underlying storage may require a system-specific configuration. Acceleration of data storage for Kubernetes is a nascent space with multiple new technology offerings, but these can require sophisticated setup and vendor integration and were not fully evaluated for this report.

*Public Cloud Managed Kubernetes Variants*

Each of the cloud providers that we deployed Kubernetes clusters on host a vast array of computing services—on-demand compute VMs, related bare metal and virtual infrastructure, specialized services for HPC, and many others. Using the public cloud Kubernetes variants does introduce some minor added cost, but we determined public cloud managed Kubernetes variants are a worthwhile tradeoff for developer time and convenience.

- *Amazon Elastic Kubernetes Service:* We utilized Amazon EKS and related computing, storage, and network services to deploy our multiple-node HPC benchmarking applications. This managed Kubernetes distribution provides a great deal of configurability necessary for obtaining the networking connection details required in public cloud Kubernetes HPC application development.

- *Google Kubernetes Engine:* In this work we utilized GKE and related computing, storage, and network services to deploy a multiple node HPC-style benchmarking application. This managed Kubernetes distribution provides a tightly integrated and convenient setup.

- *Azure Kubernetes Service:* We utilized AKS to develop workflows with the necessary configuration to deploy a managed Kubernetes cluster, but did not test the full functionality to allow HPC-style MPI applications to run cooperatively across multiple nodes. Our investigation of AKS for HPC virtual clusters is ongoing.

*Terraform + Kubernetes*

Having previously deployed containerized scientific applications with Terraform and Ansible [39] orchestration and provisioning [40], the freedom from vendor-lock-in provided by the multi-cloud features of Terraform were a very appealing option for creating a cloud cluster on which to test containers. Terraform is capable of provisioning many resources on many public and private cloud providers, including individual VMs, networking, and Kubernetes components. Thus, when choosing how to implement Kubernetes clusters in the cloud, we chose to leverage Terraform features to accomplish multi-cloud deployments. While the configuration varies by cloud provider, many public examples and modules already exist that can help ease the implementation burden.

Using Terraform, we were able to implement automated deployments in AWS EKS and GCP GKE with many customizations to support HPC workloads. Ultimately, we determined that while we could create a Kubernetes cluster with Terraform and Azure AKS, the AWS and GCP options were easier to implement with clearer documentation for the types of tasks that we were trying to accomplish. All of our Terraform and Kubernetes automation is available in the federatedcloud/kubernetes-mpi-clusters [41] GitHub repository though it is still in active development as we continue to improve our automated deployments.

## 2.2 Implementation

For each container runtime and Kubernetes, we describe here the process for implementing a scientific computing application on a variety of computing platforms from the perspective of a researcher who is unfamiliar with containerization and orchestration technologies. With the assumption that the containers and deployments will be used to conduct scientific research, we discuss additional considerations that arise. As such, some implementation choices may differ from standard industry practices, but likely follow common or best practices for the scientific computing community.

### 2.2.1 Docker

In order to run an application as a Docker container, it is necessary to create a Dockerfile which is a build script that lists all of the steps needed to prepare a container image to run an application. Dockerfiles serve as "recipes" that Docker follows when creating Docker images. Once an image has been created for an application, it can be launched as a Docker container.

Docker's online documentation [42] provides some excellent guides, reference materials, and samples to help those who are getting started with Docker. In particular, their "best practices" document [43] is a must read for new Dockerfile developers. In addition to studying Docker's documentation, the following recommendations are offered to help beginners to be more successful when creating Dockerfiles.

Dockerfiles typically contain instructions of familiar types. Often, all instructions of a type can be performed consecutively, but sometimes they may need to be interleaved for them to be successful.

Some typical instruction types are:

- *Operating System:* After the base image type is specified, the operating system must be updated so that later software installations will be successful.

- *System Settings:* These instructions might create user accounts, start and configure services, alter system-wide settings, or set environment variables. These instructions also include changing the current user and working directory.

- *Software Installation:* Scientific applications often rely on computational libraries, tools like compilers, or programming languages like Python, and their installation must be specified. If a software package cannot be installed through a package management system, it can usually be compiled from downloaded source files. Installing multiple interdependent packages often depends on finding the correct order for the commands.

- *Copying files:* The source code for an application or library, as well as any data used by the application at build time, must be downloaded or copied into the image so that it can be compiled and run.

- *Compiling:* The source code for an application may need to be compiled to produce an executable file that will run under the container's operating system.

- *Clean up:* Since it is usually desirable for a container to have the smallest footprint possible, it is a good practice to remove any files and installed software that is not necessary to run the application. This may include the application source code, the compiler used to build it, and temporary system files left over from installing software.

New Docker developers should plan to create their Dockerfiles incrementally, checking the status of their work periodically by running an interactive container from their image and inspecting it to evaluate its state. Are all of the expected software libraries and applications installed? Are the desired environment variables set to the right values? Are users and directories configured as expected? Are all copied files in the right locations and do they have the right permissions? Once developers identify commands that are useful status checks, they can include those commands in their Dockerfile to serve as tests during the build process.

This approach can be especially helpful when errors are reported during image creation. If the solution to a build error isn't obvious, comment out the lines of the Dockerfile that produce errors until the image will build. Then, run the image interactively and try to issue the failing commands *within* the container. Often, better error information is available within the container than can be seen in the Docker build output.

Another helpful technique when creating Dockerfiles is to use a VM of the same base operating system as the Docker image. This can be done either in the cloud or using a local hypervisor. One can use this VM to determine the commands (and their specific ordering) that will successfully create a working application. These are the commands that need to be included in the Dockerfile. Such a VM can also be used to compare configurations between an environment in which the application runs successfully (the VM) and a Docker image that is still being developed. In order to better simulate the container build experience, it may be necessary to recreate this VM periodically to witness again all of the changes that are happening there as commands are issued.

Be aware that as software is installed in a Linux-based Docker image, it may be necessary to issue additional commands so later commands can find the newly installed applications and libraries. These may include:

- Updating the PATH environment variable to include locations where a compiled application is located.

- Instructing the shell to "rehash" or regenerate the list of known applications that lie along the specified PATH.

- Updating the LD_LIBRARY environment variable to include locations where installed or compiled shared libraries are now located.

## 2.2.2 Singularity

In this project, we leveraged the ability to build a Docker image that can be run directly in X-Containers (more below) and converted to Singularity in order to run our application codes in different container runtimes and compared them with as much consistency as possible. This enabled us to focus our container development efforts primarily on building suitable Docker containers, with the ability to swiftly switch to the other runtimes as needed. A similar workflow would also be suitable if one already has an existing Docker container that can be converted, or if one is familiar with the construction of Docker containers and would prefer to build containers within that environment. This approach is common for HPC containers due to the disproportionate quantity of Docker container images available freely online compared to other runtimes. In some of its training materials [44], TACC follows this practice of using Docker to develop containers and using Singularity simply as a runtime to execute containers on HPC systems, accessing Singularity through the Lmod system on TACC machines (via the shell command: `module load tacc-singularity`).  Thus, it seemed natural for us to develop our Singularity containers that were run on Stampede2 in this manner as well.

**Docker to Singularity Conversion**

First, you must have a Docker image that is able to be converted to Singularity. In order to accomplish this, you will need to:

1. Create and build the Dockerfile using Singularity's recommended best practices for Dockerfiles [45]. Note that these differ from the aforementioned best practices for Dockerfiles provided by Docker.

2. Test the Docker container to ensure that the application functions correctly and all settings/configurations are as expected when running within Docker.

3. Tag the Docker image and push it to a Docker Trusted Registry (DTR).

If you are not the architect of the Docker image that you would like to use in Singularity, it is still up to you to verify that the Dockerfile followed the best practices outlined by Singularity or risk unexpected behavior or failed conversion. Most of the differences in best practices center around the variation in trust models and features between the two runtimes. In our experience, failed conversion is possible, but generally infrequent even when not all best practices are followed. However, following the suggested conventions often do not require much additional effort (if you are familiar with Docker), and a new Dockerfile can be created for conversion based on an existing one if needed.

Once you have a Docker image that is ready to be converted, Singularity provides two main mechanisms to convert a Docker image into a Singularity image:

1. The "singularity pull" command can specify a Docker image and registry.

2. The Singularity Definition file can specify a Docker image and registry in the header by bootstrapping.

Both assume that the Docker image is available on a registry that you have access to, such as a public image on Docker Hub or a private registry you can specify the connection details to [46]. Each of these options results in a rebuild of the underlying layers of the Docker container into a SIF container file. For an example of the first method, the following command will download the latest WRF container from the Cornell CAC organization's Docker Hub repository [47] and build a Singularity image named wrf_4.2.2-intel-7415915e0b8e.sif:

```
singularity pull docker://cornellcac/wrf:4.2.2-intel-7415915e0b8e
```

In the second case, for the same image being pulled from Docker Hub, the header of the definition file looks like:

```
Bootstrap: docker
From: cornellcac/wrf:4.2.2-intel-7415915e0b8e
```

If you are pulling the image from a registry other than Docker Hub, you will need to additionally specify the remote repository and configure its connection through the Singularity CLI. Due to the added features made available in the Definition file, and the ability to add the file to source control for repeatability of Singularity conversion, we highly recommend the second method for converting from Docker containers.

If you run into trouble while converting your Docker image to Singularity, Singularity has some suggested troubleshooting steps [48], and there is also a community tool for assisted conversion for users of Windows and Mac OSs [49].

**Starting from Singularity**

In other use cases, one might be interested in building a Singularity image directly, without using a Docker image, especially if one never intends to use or learn Docker (i.e., a researcher focused on working with supercomputing systems). The process is ultimately quite similar to building Docker containers, with the Singularity definition file replacing the Dockerfile. We will not describe such a process in detail here, but refer to the Singularity documentation for more information [50]. There are some benefits to this approach, such as an intuitive runscript section, the option to have integrated tests during build, and clearer organization of container metadata. However, one can also leverage these features during Docker to Singularity conversion if they choose to convert within a Singularity Definition file rather than through the "singularity pull" command. We will discuss these features in more detail in Section 2.3: "Ease of Use and Feature Comparison" on page 17.

### 2.2.3 X-Containers

As X-Containers development is ongoing, those wishing to install it should refer to information available directly from X-Containers.org [51]. The observations below were made using a proprietary installation on a private Aristotle system performed by Exotanium staff.

The instance running X-Containers is based on CentOS 7.8 and has two volumes. The first volume contains the operating system and tools for the host. The second volume is configured using Docker's Device Mapper [52] storage driver to provide a "thin pool" that stores the Docker images and containers used with X-Containers.

The thin pool volume allocates the same amount of disk space to each container. This size is configurable, with a default of 10GB. Additional configuration can be performed to specify the number of CPUs and RAM that will be allocated to each running container. In the host instance's second volume, its CPUs and its RAM should be sized appropriately to support these allocations for the expected number of containers that may run simultaneously.

Running a Docker container under X-Containers is as simple as adding the option `--runtime runxc` to the same "docker run" command that would be used to run with Docker. Many other "docker run" options will work as expected under X-Containers, though some are not fully implemented at the present time.

### 2.2.4 Kubernetes Orchestration

To explain the orchestration strategies and technologies that we used, it is useful to first describe our prototyping and development workflow for these application containers designed to run scientific software. First, a science application container is developed with small prototype tests on a single node (a single VM in the cloud). Next, a small-scale scientific test of the containerized application is performed both single node and deployed on a provisioned cluster environment to verify the multi-node capabilities of the container and cluster. Lastly, any changes to configuration or the science application container are implemented to adapt it to run scientific workflows at scale on the chosen platform.

Once the application container has been built and pushed to a public registry (Docker Hub, in our case), it can be pulled onto any compute nodes in a cluster. Using public cloud hosting services, a virtual disk image can be taken as a snapshot and used as a starting point for new VMs in a cluster [53], greatly reducing the deployment time for large clusters. Once a container has been verified as ready for adaptation to cluster scale testing, an isolated cluster environment with dedicated storage and network resources can provide useful and rapid feedback during development. Being able to scale up the same cluster configuration from development size to large-scale production runs can significantly simplify the transition as well. To provide such responsive cluster environments, we chose to provision on-demand clusters of public cloud VMs using Terraform and Kubernetes.

The basic architecture of our virtual cluster involves a launcher node running one container focused on orchestration and a group of worker nodes running a second container focused on a computational application of interest. The worker nodes communicate with each other through MPI.  Kubernetes is used to coordinate the orchestration, and Terraform is used to provision the necessary resources, although the details of what Terraform does will differ on different cloud platforms. Terraform and Kubernetes interface with each other through the Kubernetes-Alpha provider [54]. The container running on the launcher node creates a shell environment that coordinates access to the Terraform and kubectl utilities and any cloud-specific command-line utilities needed for accessing cloud resources. Terraform provisions a Virtual Private Cloud (VPC) [55] network and a subnetwork for the cluster to communicate over, and Terraform creates a cluster from a specified worker node instance type.

In addition to provisioning the basic cluster components, Terraform also supervises the construction of resources for distributed computations using MPI. To coordinate distributed computations over MPI, our system uses mpi-operator [56], a component of the Kubeflow package that supports all reduce-style distributed computations in Kubernetes. An mpi-operator configuration file—which describes the service account, the cluster role, and the cluster role binding—is used to specify the creation of those resources, which are then applied and added to the system created by Terraform. Once the cluster resources are created, the launcher container injects the relevant application input files into the container running on the worker nodes, and the MPI application is launched on the worker nodes. When the application finishes, process results are communicated back to the launcher node for retrieval from the launcher container and the resulting data files remain present on a shared NFS server data volume.

While the Terraform modules we used for each cloud flavor of managed Kubernetes distribution configuration are still under active development or provided as an alpha release, we were able to effectively utilize open source samples, documentation, and community support to achieve the necessary VM, networking, and storage settings required for HPC-style MPI computing. For more detailed discussion of the considerations and concerns with this style of deployment and how to mitigate them, see Section 2.3.4: "Kubernetes" on page 20. Other provisioning tools considered were either limited to only a subset of the necessary cloud platforms or lacked Kubernetes cluster configuration capabilities.

## 2.2.5 HPL as a Test Application

We chose an industry standard application for computational cluster throughput assessment [57], namely High Performance Linpack (HPL), as a sample application to demonstrate the functionality of cluster-based software in containers on various platforms. The HPL benchmark solves a dense system of linear equations in a parallel manner utilizing MPI, which is similar to other HPC codes and contributes to its common usage in computational cluster performance assessment. For running HPL, we chose the same medium size equation system to solve on each system without tuning to system characteristics to avoid biases introduced by underlying hardware and network properties that can vary between the computing platforms selected, such as CPU (or vCPU) core count and RAM.

*Table 1: HPL Benchmark Container Job Configuration on Various Computing Platforms*

| Platform | Container Runtime | Job Config | Node Config | Network Config | Storage Config |
|----------|-------------------|------------|-------------|----------------|----------------|
| **GCP** | Docker | Kubeflow MPI Operator | GKE Terraform Kubernetes | GKE Terraform Kubernetes | Terraform |
| **AWS** | Docker | Kubeflow MPI Operator | EKS Terraform Kubernetes | Terraform Kubernetes | Terraform, Manual Format |
| **Aristotle** | Singularity | Slurm | Slurm | Ansible OpenStack | Ansible OpenStack |
| **Aristotle** | X-Containers | Bash | OpenStack | OpenStack | OpenStack |

An HPL Docker container was developed initially and deployed in a Kubernetes cluster on GCP and AWS, with additional runs performed in Aristotle by conversion of the Docker container to Singularity and X-Containers runs of the Docker container, as illustrated in Table 1 above. In our use case, HPL serves less of the role of measuring performance, but as a way to test and demonstrate that a cluster-based application can perform in these runtimes. As such, we do not include performance metrics from our runs. Cluster configurations ranged from a single node for testing up to 4 nodes (VMs), with functional MPI communication. All container and orchestration files and scripts from these runs are open source and available on the Aristotle GitHub [58]. For technologies used to implement the orchestration of multiple node WRF runs on various platforms, see Table 2 on page 30 of Section 3.4.1: "Orchestration Methods for HPC Runs."

## 2.3 Ease of Use and Features Comparison

### 2.3.1 Docker

In the years since its release, Docker has built a reputation for a developer focused implementation and a convenient image build process that can be performed on a laptop and shipped portably elsewhere. Even for applications intended to run on a single computer, there are substantial benefits to scripting the application build into a Dockerfile. The Dockerfile's readability and simplicity make it a very convenient reference for other developers to install software from inside or outside of a container on their own systems and is thus a great candidate for storage in a source-controlled repository. Recording the application build process for posterity in the event of a software update or host migration can greatly reduce uncertainties when rebuilding an application, saving valuable developer time.

Another benefit is the natural portability of being able to run the Docker image as an application container on any Docker runtime-compatible machine including Linux, Windows, and Mac OS workstations. The portability benefits extend to other container runtimes as well such as Singularity, which has taken on the burden of supporting Docker image imports directly into its runtimes. Learning another container runtime build tool to side-step the multi-user system security concerns of Docker's root user daemon process architecture leaves room for improvement, but as mentioned previously, conversion is often successful without much developer effort. Showcasing the portability of the Docker container runtime, it has even been deployed to embedded edge computing resources and mobile robots [59].

The free hosting on Docker Hub for public images is an undeniable advantage of this runtime over many others. Private Docker images require more care and credential management to deploy, but can be hosted on a variety of services such as a free Docker registry [60], Docker Hub Professional image hosting paid service, or even served over cluster NFS or secure file transfer using Docker command line tools to save and load Docker image files, though this method is less straight-forward.

Docker image deployment compatibility is generally broad but can be limited by the CPU or vCPU kernel architecture on which the Docker image was built. This can become an issue in practice as some target public clouds offer instance type flavors running kernel architectures such as ARM [61] which are incompatible with the usual x86-64 architecture [62] system compilation. Also, some supercomputer systems which support Singularity containers imported from Docker images feature specialized process architectures or performance enhancing architecture flags best utilized by compilation on the destination system. Cross compilation and preservation of relevant libraries and binaries is possible but can create quite complicated workflows. With some care taken, Docker images can be published pre-built to scale on other platforms.

Due to its popularity among developers and a broad ecosystem of compatible tools, Docker containers are a common choice for application development on Kubernetes. With the built container providing consistency in deployment, specific parameter adjustment is possible with runtime arguments or flexible file storage attachment options during deployment. However, while Docker containers are generally well-suited for single node applications, it is more difficult to use the simple Docker developer build tools for local development of Docker images when developing HPC-style MPI applications for cluster computing across multiple VMs.

If eschewing use of Kubernetes for multiple VM deployments, cluster scaling can become cumbersome as ssh access keys and host VM identifying addresses must be provided manually to each running container. MPI implementation-specific network communication with ephemeral UDP and TCP ports can require Docker containers to use the host VM network and an on-default ssh communication port. This requirement exists in order to not consume the host VM ssh port and avoid rendering the VM unreachable. When deployed to a cluster of VMs either inside or outside of a Kubernetes distribution, file persistence such as keeping computation results must be managed, usually via a shared NFS mount. The same shared file mount can be used to fulfill MPI application expectations for a parallel file system visible to all compute nodes as in a traditional HPC system, but must be managed by the user.

Based on Docker engine, local single-machine Kubernetes distributions exist to create testing environments that mimic the destination cluster, but without careful stubbing of the destination cloud services the configuration settings may not accurately represent the fully deployed cloud cluster state and will require later modification. We discuss more Kubernetes specific concerns in Section 2.3.4: "Kubernetes" on page 20.

### 2.3.2 Singularity

One of the greatest advantages of Singularity containers is that they are able to be used on HPC resources managed by XSEDE, TACC, and other providers that have high security requirements for container runtimes on multi-user systems, as well as on cloud resources with or without an orchestration engine like Kubernetes (though the Singularity-CRI project for Kubernetes integration is not currently supported [63]). Due to the integration with Slurm and other batch schedulers, jobs utilizing Singularity containers can be queued up using HPC resource default job queues and submission methods. In the cloud, scientific applications that run using a Singularity container can be initiated, managed, and interacted with in much the same way as Docker containers via the Singularity CLI.

The origins of Singularity as an HPC-specific container technology really show through when implementing HPC applications. Singularity containers run as a user process on the host system and use the host network by default which—in addition to encouraging an entirely different security model—eliminates the need for the additional network configuration often required when using Docker (e.g., providing an extra ssh port to a deployed Docker container running a multiple node MPI application). Singularity seamlessly integrates with the host systems' root filesystem, facilitating data persistence and significantly easier data management that pervades HPC. On supercomputers, this greatly simplifies access to faster I/O while the application is running, as well as removing the need for extra configuration of storage to link the container to the host as with Docker containers. However, Singularity in the cloud still requires NFS or other storage solutions for some HPC applications in order to coordinate file operations across multiple VMs in a cluster, which is a requirement stemming from the underlying virtual cluster configuration regardless of container runtime. Once access to the data storage location is accessible on a virtual cluster, there is no need to inject input data or files into a Singularity container.

When Singularity containers are used for distributed computing, the host-level MPI libraries can be leveraged to support scaling without requiring additional MPI configuration (as required by Docker) such as hostfile injection to list the IP addresses that MPI should use to communicate between VMs in the cluster. Singularity containers can be privately distributed to all cluster nodes via a shared filesystem such as NFS or Lustre [64]. In terms of application performance, Singularity beats out Docker. When an application that has been built and run in a Singularity container is compared to similar application builds in Docker containers or on bare metal systems, applications in Singularity containers can meet or exceed the performance especially if complicated application libraries can be built with more efficient versions [65]. In common cases, container overheads, including virtualization context switches during system calls, often range to a few percent of total job runtime [66], which is often a worthwhile overhead when application and CI development time can be significantly reduced.

While it is useful to be able to build Singularity images from Docker images, this does require the extra step of importing the Docker image and converting to SIF format, which can take several minutes for a large image. Depending on the development process, multiple iterations on a large image could slow down development; otherwise, this time is usually negligible in the scheme of larger CI development.

### 2.3.3 X-Containers

For many applications, using X-Containers will be a very similar experience to using regular Docker. Existing Docker containers do not need to be modified to work with X-Containers. For many applications, only a small amount of X-Containers configuration will be needed for it to run well. This is especially true if all of the containers in use have approximately the same disk footprint and require similar amounts of CPU and RAM. In such situations, determining a single X-Containers configuration that suits all containers will be much easier. The prime benefit of using X-Containers over regular Docker containers is security. Each X-Container runs with a dedicated OS kernel, and the isolation between different X-Containers is enforced by a virtual machine hypervisor with a much smaller attack surface. As a result, a malicious or compromised container cannot easily hack other containers running on the same host.

Furthermore, X-Containers optimize application performance by turning the OS kernel into a library OS, significantly reducing the overhead of system calls.

Naturally, stronger security isolation comes with a cost, which in this case is a minor amount of overhead when containers are started, stopped or migrated. An example from our testing is that Docker's "hello world" container took 16 seconds to start and 11 seconds to stop in X-Containers, while it took 3 seconds to start and 1 second to stop under Docker. No evaluation was performed for the time required for the xSpot technology to migrate a container between instances. Beyond the starting, migrating and stopping overheads, X-Containers does not introduce any overhead to running containers.

Additionally, some planning and configuration is required for X-Containers that is not needed for Docker. And, at present, some of the less common "docker run" options may either not work for X-Containers or will require a different technical solution than is used with regular Docker.

### 2.3.4 Kubernetes

Our Kubernetes-based orchestration framework involves the interoperation of Kubernetes, Terraform, and other various tools, each of which continue to evolve at a rapid pace, including changes in APIs and configuration interfaces. One challenge, therefore, involves keeping those various tools up-to-date and in sync. Obviously, such a challenge characterizes any large collection of interdependent software components and researchers engaged in scientific computing are well-versed in navigating such interdependencies and version incompatibilities. Those researchers might be less experienced in navigating the interdependencies of tools involved in orchestration, each of which provides its own configuration interfaces and languages that extend beyond the typical world of the configure script and Makefiles that characterize many scientific codes.

Fortunately, Terraform version 1.0 was recently released, and is guaranteed to be stable for 18 months from the release date, so the challenges of updating Terraform configurations (e.g., how providers are specified and managed) are diminishing. Terraform support for Kubernetes, and options for Terraform providers for Kubernetes, are also increasing as the project stabilizes.

Additionally, our Kubernetes/Terraform framework orchestrates the deployment of two different containers, one of which runs orchestration code on the launcher node, and the other of which runs the actual computation of interest on compute nodes. Understanding this separation of concerns is important for users who wish to modify the system to run different executables. Furthermore, the container running on the compute nodes is most conveniently pulled from a repository such as Docker Hub during the configuration of a Kubernetes pod service, but this requires an additional step of pushing a new application container to a public registry as part of the overall build process if one wishes to use our framework for a new application. Since Docker Hub has become somewhat standard, this is likely only a downside for those who use proprietary software (though some solutions exist for private container hosting), or for those unable to develop their own container for computing applications.

We have developed a script to coordinate the shutdown of our Kubernetes cluster once we have finished, so that Terraform can terminate instances, delete the virtual private network, and shut down the Kubernetes engine. This method is a major benefit above typical managed Kubernetes services that may leave resources lingering at varying levels, potentially incurring unreasonable costs to researchers. It should be noted, however, that if the cluster fails to be properly constructed in the first place (as might happen during initial attempts at configuration), then the shutdown script does not always properly clean up, in which case manual cleanup of resources through the cloud GUI is required. As such, familiarity with configuring the public cloud services of interest, as well as the GUI for the cloud provider(s) being used can be very helpful when approaching Kubernetes orchestration.

When comparing the public cloud processes for Kubernetes, we found that AWS and GCP deployments of Kubernetes clusters were much easier to configure than Azure due to differences in the process of setting up appropriate provisioning account permissions for Terraform and Kubernetes. Terraform requires a service account to be set up, and good security practices would suggest not enabling more permissions to such an account than is necessary to be able to execute the deployment of resources. Unfortunately, it is not obvious how to allow access in Azure to deploy the cluster without simply granting wide open and unrestricted permissions to the service account. Additionally, we were unable to configure Azure to use the features of the Kubernetes-Alpha provider that were needed for advanced MPI applications, and are still exploring alternatives to execute large scale MPI application runs on Azure with a portable Kubernetes approach.

One of the increasingly useful benefits of Kubernetes clusters is logging and performance monitoring tools that have become widely used and supported in industry DevOps. For an HPC-style cluster deployment in Kubernetes, the initial benefits during development of containers, orchestration, and scientific workflows are not immediately apparent, so we have not spent much time implementing logging and monitoring strategies. However, once a scientific application and cluster configuration are ready to be deployed at scale, the integration of such tools can become quite useful. For a cluster designed for a single long-running job or several jobs over longer time scales (optionally scaling the cluster up and down to meet demand), the aggregated data and notifications provided by these services can be useful to ensure resources are being adequately utilized, measure application performance, and keep track of unwanted events ranging from errors in the code to failure of cloud resources. These tools can be integrated into the cluster deployments we have used, and should be seriously considered for research applications that deploy for longer periods of time in the cloud.

# 3.0 Containerization and Orchestration for WRF

## 3.1 HPC Application Considerations

While containerization technologies claim to provide many advantages to software development and deployments (i.e., reproducibility, portability, ease of use), HPC applications can present unique challenges that alter the possible benefits and disadvantages of leveraging a container. Some of these challenges include:

- Increased complexity of software installation, configuration, and/or compilation.

- Specialized or highly tuned versions of the software designed for specific systems or hardware with assumptions about underlying access to such systems.

- An intersection of many performance bottlenecks (such as memory-bound, I/O bound, and/or compute-bound) usually at a higher level of concern than other application types.

- Frequent or drastic revisions and updates to accommodate the latest hardware, better algorithms (for science or performance), and even much needed usability changes or bug fixes.

Therefore, when researchers are considering whether a container will ultimately benefit their work, increase productivity, or take advantage of the aforementioned advantages potentially provided by container technologies, they must also find a balance with these challenges and the added effort of CI development. Since these hurdles tend to exist for HPC applications regardless of how they are packaged, the benefits of containerizing are often an attractive option regardless.

The development mindset of containerizing for HPC typically differs from standard container development in a few ways. The underlying architecture of the system must be considered during build and deployment in order to accommodate software configuration and compilation processes that may assume a distributed memory architecture, or may have significantly different performance characteristics depending on hardware. Application performance outside of a container cannot necessarily be assumed to be comparable to performance inside a cluster of containers, as performance comparisons of containerized workflows are an ongoing research effort [67][68][69][70][71][72][73].

Furthermore, one cannot assume that the container will be standalone. This is because the container will likely be replicated and networked together with other nodes in a cluster (virtual or on a supercomputer) in order to reach a scale to support scientific computing needs. HPC containers require configuration for MPI communication that will be supported by the system that they are deployed on. Input and output data processes are expected to happen during container runs, so paths and configurations must accommodate this, as well as I/O speeds for storage connections to containers.

Despite the complexities of this development process, many HPC containers have been developed and more HPC applications are being containerized. Once a reasonably functional container image is available for an HPC application—especially if the image is made publicly available—there is frequently an uptake in adoption of containerization techniques and tools for significant portions of the users of the software. Exceptions to this trend occur when the available container images have not been shown to meet high enough performance standards for the application, or other cyberinfrastructure challenges, such as orchestration, impede the possibility for wider adoption.

As computational scientists, research software engineers, and others who develop cyberinfrastructure specifically for HPC applications have increased activity in the area of HPC container development and improved software tools to accommodate containerized workflows, many studies have been done to measure and improve the performance characteristics of such methods.

## 3.2 WRF Use Case Description

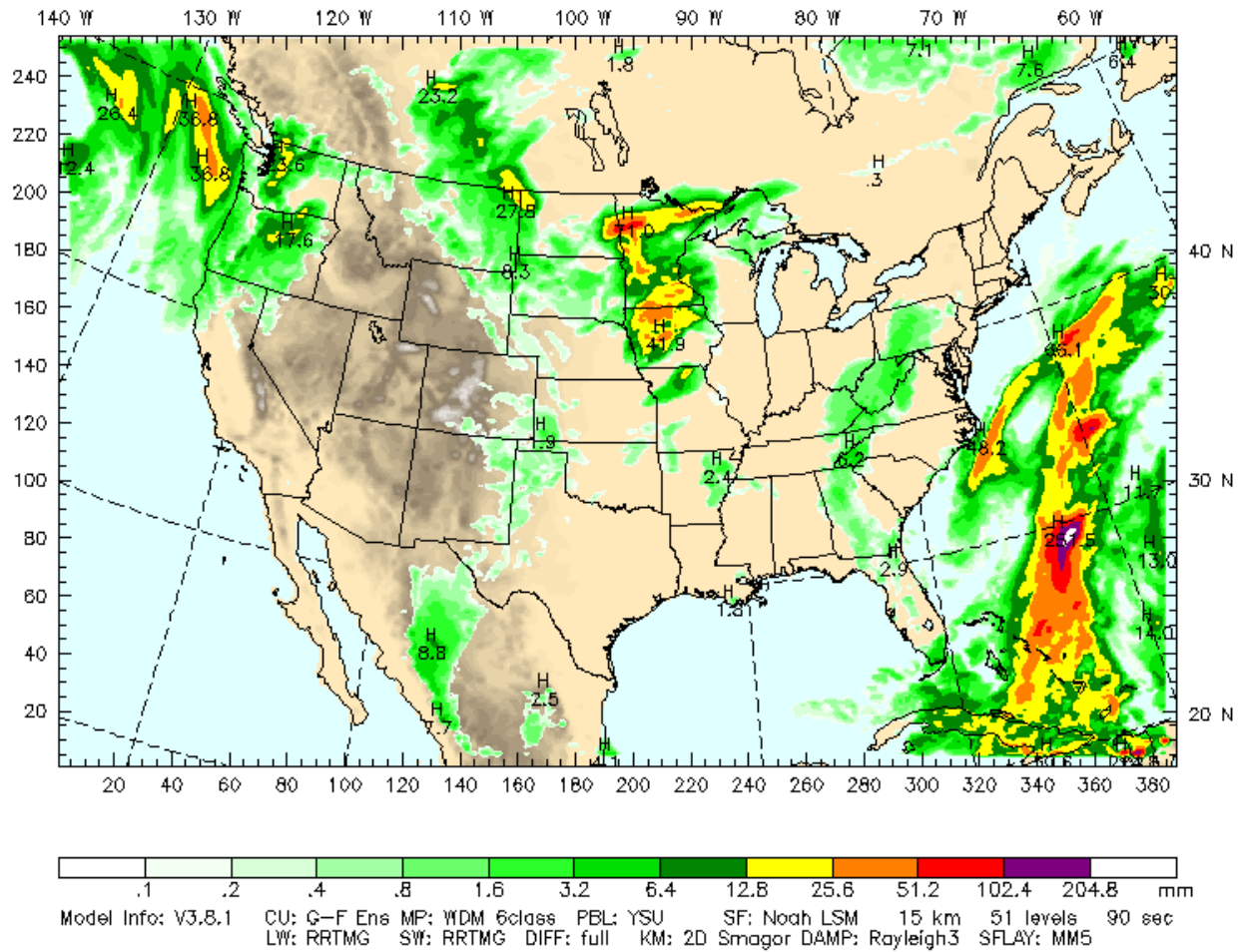

*Figure 1: An example of model domain used by NCAR forecasting group*

The Weather Research and Forecasting Model (WRF) [74] is a large open source software framework consisting of many components to enable the forecasting and advanced research simulation needs of a wide community within atmospheric sciences. WRF weather simulation is used operationally by the National Weather Service to develop hourly weather predictions for many people across the United States, along with a network of other forecasting services. The WRF model sees active usage from many other business and research organizations, and is featured frequently in computational performance assessments of new, large clusters [75]. The total size of the WRF user community encompasses at least 48,000 registered cumulative users across 160 countries [76].

As part of the Aristotle project, Cornell atmospheric scientists in the Sara C. Pryor [77] research group and their international collaborators have already benefited from our development of WRF version 3 containers and cloud deployments that allowed for increased productivity and efficiency, portability to on-demand cloud resources, and reduced wait times to start large-scale jobs compared to some backlogged, queue-bound HPC systems. This style of deployment adds the advantage of being able to test and debug science cases in real time, which reduces the time taken to complete production runs. WRF is often prone to segmentation faults due to memory demands and grid decomposition, which can sometimes take a number of days to resolve. Having access to a system that allows the scientists to run different configurations and model setting changes in real time means that they can often debug a science case faster than what it would normally take on HPC systems. Despite regular security patching on the host systems of HPC and virtual clusters, Docker containers with no exposed networking ports can be maintained at a stable operating system version for consistent software support. As a result of our support, Aristotle scientists have performed extensive simulations [78] over many months (or years) using the same container, and this has allowed for outstanding consistency in the science runs.

### WRF Offshore Wind Farm Simulations

Our research collaborators had a specific interest in assessing the impact of wind turbine placement within offshore wind farms on sustainable energy production [79]. This required moving up to WRF version 4.2.2, switching to Intel compilers within the container, adding a proprietary software package for explicit wake parameterization (EWP) [80], and scaling up the resolution of WRF model simulations (including nested domains). This science use case presented a picturesque opportunity to highlight what ongoing scientific research looks like with containers for HPC applications. Motivated by this science use case, we tested our WRF containers and orchestrated runs with CONUS benchmarks.

### Continental US (CONUS) Benchmarks for WRF

When running WRF, the user has a vast array of choices and options that ultimately can cause the numerical results of a simulation to vary. Even considering identical initial conditions, one may obtain different results on different systems. Additionally, performance is an important consideration for WRF users due to the high computational cost that can be incurred for larger simulations and forecasts. Benchmarking is a common practice to evaluate or compare WRF runs between different systems, compilers, or choice of dependencies and library options.

We chose to use the WRF version 4.2.2 CONUS benchmarks provided by NCAR to evaluate our bare metal and containerized WRF deployments. The WRF Users page [81] has a large collection of resources for WRF users, including some useful benchmarks. The WRF v4.2.2 Benchmark Cases [82] page is where you can obtain background information on the benchmarks, links to the input data used to simulate the benchmarks, compressed data archives, and some information on the simulations performed to generate the benchmarks. The CONUS benchmarks are available in two different grid point resolutions—CONUS 12km and CONUS 2.5km. The lower resolution CONUS 12km benchmark has a faster run time and is most suitable for testing initial single node setup and workflow with realistic data input. The higher resolution grid simulation CONUS 2.5km requires more compute and therefore is more suitable for exercising multi-node functionality on parallel compute cluster resources (virtual or bare metal HPC).

## 3.3 WRF Container Implementation

### 3.3.1 Why Containers

Taking into account the above challenges and opportunities presented by combining an HPC application like WRF with the software development methodologies of container deployments, we find that our research collaborators are satisfied with the resulting convenience and acceleration of scientific results. Containers on HPC systems are well-supported and scalable, meaning a WRF Docker container developed once for cloud deployments can also be converted to Singularity and reused on a variety of HPC systems. These containers further enable application run bootstrapping and faster code sharing within teams and with the wider community, once a container has been published and shared. However, initial creation of a WRF container and automation of application runs—whether benchmarks or scientific simulations—can be quite difficult and time intensive even for domain experts due to a number of complications.

WRF is a notoriously difficult application to configure and compile. One must choose many software aspects from the desired WRF major and minor version, any relevant recent source code bug fixes, and applicable custom physics model additions such as hydrologic effects or high-resolution wind turbine wakes. WRF users are often more accustomed to having no portability of WRF builds, and expect to have to reconfigure and recompile on every system they intend to perform runs on. This is evidenced in the public WRF v4.0.3 container example [83] which does not even contain a compiled version of WRF, but rather includes a tutorial to walk users through compiling themselves for their target system (effectively diminishing the benefits of leveraging a container). Some of the motivation to rebuild is due to performance considerations, but it is even common to recompile on the same system for different types of science use cases.

There is no standard path for setting up WRF software, though common examples and tutorials exist for compiling the Advanced Research WRF (ARW) core and WRF Preprocessing System (WPS) packages [84], leading many WRF researchers at a loss to solve system-specific or package-specific compilation issues that arise when switching systems or simulation types. The WRF User Forums [85] are a graveyard of many such unanswered compilation issues from users. Additionally, there are several dependencies of WRF that must be version-matched, configured, and compiled correctly in order for WRF to perform correctly after compilation (even if compilation did not yield any errors). Despite this necessary mapping of dependencies, there is no clear advice on which versions to use for your scenario, forcing users to rely on sysadmins and support staff to help them narrow down the right software for their ecosystem and desired simulations.

Still within the realm of compilation difficulties is the choice of compiler. Frequently, users will compile with GNU during development due to its free availability and easy package installation via Linux package managers. But on large-scale HPC systems, Intel CPUs are common, and the Intel compiler can provide noticeable performance improvements for WRF simulations. On these systems, the underlying hardware and compiler choice (Intel vs. GNU) consequently affect which MPI library one can use, further complicating the process of managing a working WRF implementation out of possible configurations of the WRF software, compilers, dependency versions, and libraries. In the case of our aforementioned Stampede2 test environment, the Intel Omni-Path interconnect limits which MPI libraries can be used for multi-node MPI jobs, making the Intel compiler a more reasonable choice. Therefore, it is more common to use the Intel compiler on HPC systems and for large-scale WRF simulations.

Adding in the complexities of compiling proprietary add-on software packages, as is the case for the Pryor group, can cause further portability and compilation issues that having a single standard configuration and compiling within a portable container can help to alleviate. If one is planning to deploy on multiple systems, it can be helpful to try to match common underlying hardware, and thus have one WRF container that will work on many systems with the same build. Since each of our test environments used Intel nodes (VMs or HPC system nodes) in the Haswell to Skylake line of processors, we were able to create one portable container build for the benchmarks case, and extend this container to include the proprietary software.

WRF data management presents additional challenges that must be faced. One must obtain reference geographical and time varying atmospheric input data in correct formats for the chosen simulation WRF kind, version, and custom physics models in use. Upon starting a WRF simulation, a researcher must successfully configure more than 100 application variables in the namelist file which control a plethora of detailed simulation options. System requirements must be accounted for, such as managing data storage working directory linking to compiled executables and input data, as well as predicting maximum resource needs for node memory, networked file system disk usage, and network congestion stalls not to exceed MPI latency thresholds. Long running jobs with multiple node processes executing are best supervised or instrumented to alert for fast debugging upon any error that may stem from the above choices. Using a container to package up all of the above with well-attributed scripts destined for runs on a particular system can greatly reduce the complexity of deploying multiple jobs on a system or multiple systems (especially if automation is generalized).

### 3.3.2 Our Implementation Strategy

Following our container development process laid out in Section 2.0, we first developed a few different containers to test singlenode and deploy on various platforms. We initially created a container for GNU-compiled WRF 3.8.1 [86] that had some patches to improve code functionality for the now unsupported version of WRF. Our collaborators initially were working with this version and wanted to move from compiling with GNU to Intel, so we also compiled WRF bare-metal on Stampede2 with both GNU and Intel for testing and comparison purposes (container build files included in the linked GitHub repository). During testing, the proprietary science software package failed to accomplish nested domain runs on real data with either of these compilations. Following extensive troubleshooting and testing, we determined that the failures were specific to the old WRF version, and instead moved to WRF version 4.2.2 compiled with Intel. The move was also a useful upgrade for our collaborators because they previously had issues getting a stable Intel compilation with their software combination.

#### Compiling WRF

We prepared bare-metal implementations of WRF 3.8.1 and 4.2.2 compiled with Intel 18.0.2 on TACC's Stampede2 supercomputer as a first step to producing a standard Intel compilation process that could be containerized [87]. We tailored this implementation to the Intel Skylake nodes knowing that we would deploy on cloud nodes with similar hardware (in the same line of Intel processors). Using the Lmod system, we were able to load modules for most of the dependencies of WRF with "module load" commands in the configure and compile scripts, significantly decreasing the burden compared to compiling the dependencies ourselves, which was later required for container implementation. Unfortunately, the aforementioned version-mapping confusion of compiler, MPI library, and dependencies made for a non-trivial "trial-and-error" journey that ultimately resulted in a TACC consulting ticket to resolve dependencies and get a successful compile.

The process of getting a successful WRF compile on a new system should not be underestimated, let alone an optimal and repeatable solution that can then be containerized. The WRF compilation errors are often buried thousands of lines deep in an output file, and not always intelligible to even a competent computational researcher. Several configurations in the older version of WRF, and even some in the new version, had to be injected into the configuration file to ensure a proper compile. The WRF user forums, past experience, and many attempts were ultimately necessary to narrow down issues enough to even pose the right questions about the implementation to TACC sysadmins familiar enough to help. In fact, the TACC-suggested solution involved an extensive injection into the configuration file that would have been hard to come up with on one's own. For new WRF users, this is exactly the nightmare that a container could help them avoid; and so, we endured.

Upon finding the appropriate combination of dependencies for WRF 3.8.1 bare metal on Stampede2, the transition to WRF 4.2.2 compiled with Intel was surprisingly smooth. We loaded the TACC default modules to get started, some of which—such as git—were required for our workflow. Then `intel/18.0.2`, `impi/18.0.2`, `netcdf/4.6.2`, `pnetcdf/1.11.0`, and `phdf5/1.10.4` were all loaded within our scripts during configuration, compilation, and any subsequent runs to ensure a standard environment with appropriate dependencies. Leveraging the NetCDF4, parallel NetCDF4, and parallel HDF5 libraries provided as modules by TACC meant that the dependencies were compiled with Intel and worked with Intel MPI, as tested by TACC sysadmins to ensure optimal performance of codes using them.

**Bare-Metal Testing**

Once we were able to successfully demonstrate simple testing runs with our functional bare-metal compile of WRF 4.2.2, including successfully running with nested domains, we were confident to start deploying runs on Stampede2. We scripted the data management, environment configuration, and job initiation of the CONUS 12km and 2.5km benchmark runs, singlenode and multi-node respectively. All scripts and files used that were not already provided by NCAR are available in the GitHub repository federatedcloud/WRFv4-Benchmarking [88]. These initial job scripts were the starting point for automating the following containerized runs of the same benchmarks, and served as a useful point of comparison for performance characterization and functionality. When troubleshooting containerized runs, it was very useful to have a bare-metal implementation to return to and perform sanity checks with, especially when trying to match dependencies.

### 3.3.3 Docker

**Intel Compiler and MPI Library**

In prior Aristotle work, we deployed various Docker containers all WRF 3.8.1 or older compiled with GNU and using the OpenMPI library. The Docker container developed for this work was WRF 4.2.2 compiled with Intel and using the Intel MPI library. For this version, we leveraged the Intel oneAPI HPC Toolkit [89] via the Intel-supported container images recently made publicly available [90] including the classic Intel C/C++ and Fortran compilers, as well as the Intel MPI library. Because these compilers are closed source and were previously not distributable in containers, we are not aware of any other WRF containers that are distributable with Intel prior to our work. We used the Ubuntu 18.04 Intel container as a base image, configured the environment to use the Intel compilers and MPI library, and then added our dependencies and software. Our container build scripts are available in the Docker directory [91] of our federatedcloud/WRFv4-Benchmarking GitHub repository for reproducibility, and the `cornellcac/wrf` public image has been made available on our Docker Hub organization [92] at the `4.2.2-intel-7415915e0b8e` tag [93] (the hash in the tag refers to the Intel base image hash). The container image is rather large (~25GB uncompressed) due to the base Intel image containing several extra Intel components (starting at ~23 GB uncompressed), and it is in our plan to create an image of reduced size for quicker deployments.

**Compiling Dependencies**

Despite the convenience of the Intel compilers and libraries built into the Intel container images, they were not configured to be used by default. Several environment variables and configuration steps that are often already configured for users of HPC systems had to be determined and implemented by us within the container. Furthermore, dependencies had to be compiled with the Intel compiler and configured correctly by us instead of HPC center sysadmins for the target system. For researchers unfamiliar with the lower-level details of software and intricate HPC compilations, this could cause containers to become an untenable solution, despite the possible benefits. Since we were making an effort to match the Stampede2 bare-metal compile for easy comparison, we again reached out to TACC support to understand how certain WRF dependencies were compiled in order to be able to develop the container successfully.

Despite previous experience installing WRF and dependencies in Docker containers, compiling these dependencies with Intel was new territory. When using the GNU compiler with OpenMPI, the compiler itself, MPI library, and several of WRF's version-matched dependencies can all be easily installed via a Linux package manager. Often the package manager can be helpful in finding the appropriate version of dependencies to use as well, due to the login built into the packaging. However, dependencies compiled with Intel must be accomplished by the user, and the process is not nearly as straightforward.

For example, one major impediment proved to be the NetCDF4 compilation process. When running bare metal, loading 2 modules (`netcdf/4.6.2` and `pnetcdf/1.11.0`), was all that was required to enable full features of this software for our WRF 4.2.2 Intel compile. However, in the container, the full build process involved downloading and compiling pnetCDF 1.11.0, netCDF-C 4.6.2, and netCDF-Fortran 4.5.3 each configured separately with their own compiler flags that differed slightly, troubleshooting the compile during container build iterations, and, finally, enabling NetCDF4 via an environment variable. The process was time consuming to get correct, and sometimes a problem with the compilation was not discovered until attempting a WRF simulation within the finished container, resulting in another time consuming rebuild of the container after finding the fix.

If one is not using Intel, some of these difficulties are easily overcome, but the flexibility to control the compilation of dependencies can also be a benefit for some researchers. Ultimately, the finished container with functional dependencies, libraries, and WRF outweighed the disadvantages of a difficult compilation process simply because having the container means we do not have to repeat the build. All future WRF deployments can begin from a better starting point in the finished container.

**Configuring and Compiling WRF in Docker**

After the dependencies of WRF were compiled within the container, the WRF 4.2.2 Intel compilation in the container was rather smooth because we could follow the same steps as performed on Stampede2 due to dependency version matching. It should be noted that we aimed to provide a resulting container in which the WRF software itself could be recompiled if necessary. The resulting container image already has a compiled version of WRF 4.2.2, but if one wanted to change the compile option, for example, one would only need to re-run the configure and compile steps documented in the Dockerfile with the intended modifications.

### 3.3.4 Singularity

There are many possible paths we could have taken to create Singularity containers for WRF, but we have found that converting our existing functional Docker containers is often the simplest solution assuming the container works with available libraries and configuration on your target system. For our GNU-compiled WRF Docker containers with OpenMPI, converting them to Singularity was a seamless transition. Once converted and pulled to a target HPC machine, it was only a matter of loading the correct module (`tacc-singularity/3.7.2` for Stampede2) and writing a submission script to test singlenode applications. For multi-node applications, however, one must take into account the MPI configuration, and Stampede2 does not have any configuration of OpenMPI that works multi-node due to the Intel Omni-Path interconnect hardware [94].

Singularity allows MPI communication in 2 main ways [95]:

1. Hybrid model - the container has an MPI version that matches at least the major version of the MPI version on the host (can be newer), MPI processes on the host are connected to the MPI within the container, and the processes work together to run the MPI job.

2. Bind model - the container does not have any MPI implementation and the host MPI is mounted (via a "bind" flag issued at runtime) into the container.

If you are using the Bind model approach, your Singularity container would not contain an MPI implementation at all, making it distinctly different from our Docker containers. This approach did not appeal to us because we desired to use a Docker container and Singularity container that had the same contents, down to the version number, if possible.

If you are using the Hybrid model, however, you have to match the host MPI implementation, and this is exactly where we ran into some trouble running Singularity containers multi-node on Stampede2. Initially, we explored possible paths to running our existing GNU and OpenMPI WRF 3.8.1 Docker container (once converted to Singularity) multi-node on Stampede2, but eventually accepted that we would need to move to Intel's MPI library in order to enable multi-node.

Originally, we expected that the Intel OneAPI HPC Toolkit container images—which we used as a base image in our Docker container— could provide this convenient matching implementation. Since the Intel compiler and MPI library in the Intel base images were newer than those on Stampede2, we expected the Hybrid model to work for us, even though the versions were slightly different. After the painstaking construction of the Docker container with Intel-compiled WRF and dependencies that matched the versions used in our bare-metal runs, as well as the Intel MPI library, we converted to Singularity to test on Stampede2. Immediately we ran into errors that made it clear this process would not be so simple.

The Intel Omni-Path hardware on Stampede2 actually requires extra drivers and software to be installed within the container as well, regardless of whether you are using the Hybrid or Bind model approaches. Consequently, it is not recommended that you create your own base image with Intel MPI, but rather use TACC's pre-configured base images as a starting point [96]. Unfortunately, this significantly decreases the portability of the Singularity container images created, which TACC openly admits in their container training about these images [97], but it is the cost of the specialized hardware that Stampede2 has. We were not able to construct a portable version of a Singularity container for WRF 4.2.2 that would work on Stampede2 *and* any other system. Additionally, the version of compiler and MPI library that worked with the dependencies we used to compile WRF bare-metal were not available in any of the base images provided by TACC for Stampede2.

The hardware-specific dependencies needed within the container is a portability issue specific to Stampede2; it is also characteristic of how different the HPC landscape is compared to cloud. HPC systems are not designed to be standardized or similar to one another. One cannot assume that an application configured and tuned for one HPC system will work the same and have the same performance on another due to the differences in system configuration, hardware, available software, etc. Similarly, one can generally not assume that a container developed for one HPC system will be generally portable to other HPC systems without some changes. This was possibly the most disappointing realization in this study for the state of HPC containers, but leaves many possible avenues open for improvement in the HPC space to provide portability of application containers across systems.

### 3.3.5 X-Containers

Because X-Containers were not on a track for the scientific needs of the WRF users that we were collaborating with, we did not perform full science runs or CONUS benchmark runs within this runtime. We did import the WRF 4.2.2 Intel Docker image into X-Containers running on a VM in Red Cloud to perform some basic functionality tests. Due to the high demand of resources required for WRF runs, we stopped there, but expect that the container image would have behaved similarly to the Docker image in Red Cloud.

## 3.4 Application Container Orchestration

### 3.4.1 Orchestration Methods for HPC Runs

Now armed with a selection of built and tested WRF application containers and a bevy of proven cluster deployment technologies for those containers, we sought to validate WRF application runtime performance at a moderate distributed scale. Note, we described many technical aspects of container development and multiple node capable MPI application orchestration including Kubernetes deployment using our test application HPL in Section 2.0: "Container Technology Comparison," as well as the technologies and system platforms chosen. Here, we describe the orchestration methods used to deploy WRF simulations at various scales on a variety of platforms and specific implementation concerns for running similar compute and data intensive HPC-style applications on public cloud Kubernetes resources.

**Choice of WRF Orchestration Technologies**

We chose to run WRF on platforms where we could quickly develop both scientific and performance comparison results. Our private cloud Aristotle provides a great level of control over system settings, enabling us to prototype quickly at a low cost and run long, consistent simulations. We chose Stampede2 for our bare metal comparison baseline against the WRF container implementation because of its substantial single node processing power and the availability of the Singularity container runtime and MPI performance profiling tools. We chose AWS for the comparison to Stampede2 because it provides a server instance size flavor with processor hardware that closely resembles Stampede2 nodes. In addition, AWS regularly introduces new features and services for storage acceleration, same rack placement, and other options that warrant future HPC application performance tuning investigations. Other public clouds, with their myriad offerings, may be just as suitable for performance comparisons.

WRF commonly runs on bare metal servers to provide close hardware access and to enable the highest performance from compiler optimization, while avoiding virtualization overheads. Singularity is a container runtime designed for use on HPC systems and naturally integrates with cluster job schedulers. Due to MPI library compatibility requirements between containers and host systems discussed in Section 3.3: "WRF Container Implementation," we decided to focus on single node rather than multi-node usage for this benchmark test. On public cloud resources, we leveraged Kubernetes software's quick scaling deployments together with the convenience of building the Docker container runtime. Converting Docker images to Singularity runtime images was also a useful feature for fast portability between systems.

Resource provisioning on public cloud was undertaken with tools selected for maximum reusability on multiple cloud platforms. OpenStack is the default means of managing cloud resources on Aristotle, allowing for convenient dashboard and command line configuration, as well as portability to XSEDE resources running OpenStack. Terraform and Kubernetes allow for powerful configuration of on-demand cluster infrastructure and container service orchestration by design. Table 2 summarizes the technologies we used to successfully demonstrate WRF orchestration.

*Table 2: WRF Container and Orchestration technologies used across platforms*

| Platform | Container Runtime | Job(s) | Cluster Orchestration | Network Config | Storage Config |
|---|---|---|---|---|---|
| Stampede2 | Bare Metal | CONUS Benchmarks | Slurm | - | - |
| Stampede2 | Singularity | CONUS Benchmarks | Slurm | - | - |
| AWS | Docker | CONUS Benchmarks | Terraform + Kubernetes EKS, Kubeflow MPI Operator | Terraform + Kubernetes | Manual NSF Mount |
| Aristotle | Docker | CONUS Benchmarks | OpenStack, Bash Scripts | OpenStack | OpenStack |
| Aristotle | Docker | WRP EWP Science Simulation | OpenStack, Bash Scripts | OpenStack | OpenStack |

### 3.4.2 Public Cloud Kubernetes WRF Application Orchestration

**Complexities of Orchestrating WRF**

We uncovered many implementation challenges when running compute and data intensive HPC-style applications on public cloud Kubernetes. We are sharing our implementation code in a public repository [98]. Our code reflects the current state of experimental support for MPI applications managed by Terraform Kubernetes provider constructs that allow for automatic node count scaling and cloud portability. The software-based resource provisioning can best be attempted on AWS at the present time; it also works with Google Cloud Platform with MPI applications. Conversion to other cloud platforms should be possible with extensive changes to a platform-specific Terraform Kubernetes provider or other resource configurations. Familiarity with Kubernetes software concepts, resource provisioning on the desired cloud, and debugging parallel computing applications are recommended.

**Data Management**

In contrast to our MPI test application HPL, a data-intensive application like WRF reads and writes gigabytes to terabytes of data and requires a parallel file system that all nodes can access concurrently. WRF is built for HPC system architectures that commonly provide an accelerated parallel file storage system attached via a fast network interconnect.

We tried multiple parallel file systems and data storage options offered by AWS, but ended up implementing a standard NFS server in a Kubernetes cluster pod. A large EBS disk volume was manually formatted with an XFS file system to be attached to the NFS pod host instance. Other options such as a Lustre offering, EBS multi-attach, or S3FS fuse mounting of an S3 bucket either exhibited overly complex configuration needs to tie into our provisioning or did not meet our needs for terabyte scale, fully parallel file system access. Other configurations and offerings that recently became available were not evaluated due to integration complexity.

**Networking Configuration**

With public cloud resources, setting up networking configurations to appropriately open security group ports that allow communication to and among nodes is a necessary step with some iteration required. We encourage the reader to consult our implementation [99] or the Terraform AWS EKS guide [100] as a good starting point for creating a list of necessary networking ports. Once usual defaults like ssh and external network access to download internet resources are opened, discovery and communication within the cluster is handled automatically by Kubernetes. Kubeflow MPI Operator handles MPI host listing through pod discovery and port mapping automatically as well.

**Cluster Node Configuration and Deployment**

We chose an AWS instance type that was comparable to Stampede2's processor type and core count. Initial provisioning testing was performed with smaller instance sizes and server counts to establish basic configuration and MPI networking capability using our HPL test application. Use of dedicated, bare metal, or larger public cloud instance hardware types either single node or in Kubernetes provides the possibility of faster inter-process communication within a node and can improve performance; that analysis was beyond the scope of our current work.

Deployed nodes required several modifications at boot time to facilitate Intel MPI multi-node communication among Docker containers orchestrated by Kubernetes; this was accomplished by a combination of Terraform volume resources and AWS user data scripting. In particular, Docker engine stack memory limits were set to unlimited in a user data boot script and a 100GB disk was mounted onto worker node/dev/shm to provide for sufficient shared memory use by MPI.

We followed the typical Kubeflow MPI Operator cluster layout of a launcher node and multiple worker nodes to perform most computational work. MPI slots were chosen to match the virtual core count of the instances used without attempts at fine tuning. Individual containers occupied all available resources of each worker node. Worker nodes were provisioned in an auto scaling group to allow for easy expansion of the cluster to large node counts between runs. To run, the simulation can either be scripted into a Kubeflow MPI Operator MPI job within Terraform or the user can ssh into the launcher node and submit the job using the command line.

## 3.5 Performance

### 3.5.1 Performance Characterization

For this study, we generally focused on capturing high-level performance statistics, such as aggregate application run times, without doing detailed dives. In addition to application run times, one issue of potential concern in comparing orchestration-based applications in the cloud with queue-submitted applications on HPC resources is the time to application start-up. On a scheduled HPC system, applications might sit in a queue for some amount of time before actually starting. With an orchestration-based application in the cloud, one must account for the time required to spin up the various resources that are required. While the spin-up time on a cloud resource will generally be fixed from run to run (but can of course vary depending on the orchestration strategy), the queue wait time on a shared resource can vary more broadly, depending on the number of other jobs currently running or submitted.

Collecting application run times can be accomplished using the Linux `time` command—similar to typical application runtime logging—which will be available within a Docker or a Singularity container. Methods for collecting more detailed performance statistics will depend on what sort of information is needed. For applications built using the Intel suite of compilers, the Intel VTune suite of applications can generate detailed reports addressing various aspects of application performance [101]. For MPI-based applications such as we are running here, there is additional guidance on how to use VTune to characterize performance [102] [103]. On HPC systems at TACC, such as Stampede2, the REMORA (REsource MOnitoring for Remote Applications) tool available through the REMORA Lmod module can be used to characterize application performance [104]. For simpler sets of statistics for jobs run on HPC systems using Slurm, the `sacct` application (part of the Slurm suite) can be used to query jobs after they have been run to assess job submission times, start times, end times, and so on.

It should be noted that any additional tools required for performance characterization of the application itself must be accessible within a container that is being used to run the application. For Intel-compiled codes, the aforementioned Intel oneAPI container used as a base image includes the VTune suite, although the much slimmer oneapi-runtime container that can be used to support Intel-based codes does not contain the VTune suite. Therefore, if one were to use the streamlined oneapi-runtime container and use VTune to characterize application performance, one would need to install VTune additionally into the application container.

Multi-node performance characterization of containerized applications can become more complicated when considering networking between nodes, especially orchestrated deployments in the cloud. Because public cloud resources are often shared among several users, one cannot assume that they have access to a full VM's capabilities. To truly capture multi-node application performance, one would need to perform multiple runs at varying times to characterize the effects of other cloud users on the cluster network communications. Additionally, the application memory and CPU utilization could be affected by the density and activity of other users on the same underlying hardware. Since this study of container runtimes does not delve into these cloud computing performance characterizations, we mention these for users interested in cloud deployments with containers to keep in mind.

### 3.5.2 Results

**CONUS Benchmark Runs**

Beyond the single and multiple node tests to verify functionality of the containers and orchestration, we ran a variety of experiments to characterize overall performance. Runtimes were captured for the CONUS 12km and 2.5km benchmarks on a variety of platforms and configurations. We present a sample of these runs in Table 3 to give some insight into comparison of performance of our containerized runs in the cloud (in AWS in this case) as compared to bare-metal on Stampede2.

*Table 3: Sample WRF CONUS Runs*

| Platform | Container Type | Virtual-ized | Node/VM Type | CPUs | RAM | Job Nodes | Application | Runtime |
|---|---|---|---|---|---|---|---|---|
| Stampede2 | Bare-Metal | No | SKX | 48 | 192 GB | 1 | WRF CONUS 12km | 1m54.621s |
| | | | | | | 4 | WRF CONUS 2.5km | 21m9.043s |
| AWS | Docker | Yes | c5.9xlarge | 18 (36 vCPUs) | 72 GB | 1 | WRF CONUS 12km | 2m29.528s |
| | | | m5n.24xlarge | 48 (96 vCPUs) | 384 GB | 4 | WRF CONUS 2.5km | 21m5.963s |

The CONUS 12km benchmark runs were performed on a single node for each system, and only 36 cores of the 48 cores on the Stampede2 SKX node were utilized in the Slurm submission script. The c5.9xlarge VM instance in AWS is therefore approximately equivalent in CPUs (vCPUs for AWS) to the Stampede2 job, though there is much less memory available on these VMs compared to the SKX nodes. The CONUS 2.5km benchmark runs were multi-node, specifically 4 nodes on each system. We used the SKX nodes on Stampede2 again, and in this case, we tested the m5n.24xlarge VM instance type on AWS for this run, which is arguably larger than an SKX node in vCPUs and RAM, but the physical cores match the SKX nodes. The c5.9xlarge and m5n.24xlarge VMs on AWS have a "1st or 2nd generation" Intel Xeon Platinum 8000 series processor (Skylake-SP or Cascade Lake) with a sustained all core Turbo CPU clock speed of up to 3.1 GHz and 3.6 GHz respectively [105] [106]. Each of these VMs is therefore an Intel processor roughly equivalent to those on the SKX nodes of Stampede2.

As one would expect due to no virtualization or containerization, the bare-metal run of the CONUS 12km benchmark ran faster. Interestingly, the CONUS 2.5km run on AWS Kubernetes cluster deployment of WRF 4.2.2 Intel in Docker out-performed the bare-metal Stampede2 run. We would expect the bare-metal run to perform faster, as in the CONUS 12km case, but there are a couple possible reasons to explain this discrepancy. The m5n.24xlarge instance type has 48 physical cores like the SKX nodes, but 95 vCPUs, which could ultimately accommodate more compute if competition from other users on the hardware is low. Also, there is significantly more memory available on these VMs than on Stampede2, which could speed up computations where memory may be a bottleneck.

In our testing, we noticed that memory availability did affect WRF runtime results. When deployed on a small VM, WRF would fail very early in an attempted CONUS benchmark run, and large memory allocations to begin the run could not be completed. After increasing the VM's size—which includes increasing vCPUs and available memory— WRF simulations began smoothly with large amounts of memory allocated to the Docker containers' shared memory to enable MPI process communications. While we saw memory usage increase, the increased compute available did not seem to increase CPU utilization as dramatically. Consequently, we expect that the difference between runs for the CONUS 2.5km benchmark could be due to larger memory available on the AWS VM. Certainly, more runtime and performance information from different VM sizes as well as repeated runs on these (and possibly other systems) could yield interesting insights.

While we have not included specific numbers in the above table, we did also see some expected differences in the deployment time of jobs on the Stampede2 system vs. the cluster spin-up and start time of jobs in AWS. The AWS Kubernetes cluster typically takes around 10 minutes from initiation until it can begin running jobs. If one is running multiple tests, experiments, or long simulations, then this is a one-time cost. All future jobs can be started immediately. However, on Stampede2, we noticed even the 1 to 4 node CONUS jobs (relatively small for the system) sat in the queue for times ranging from about 2 to 40 minutes. The size of the job and the time of day submitted can significantly affect the amount of time a job will wait in the queue. Similar to AWS, once a container or cluster is deployed in Aristotle, consequent jobs can be run immediately without any wait time; this is a fact our collaborators were quick to point out as a major benefit of this work.

## Science Runs

After verification and experimentation with the WRF 4.2.2 containers for CONUS benchmark runs, we extended the Docker container to enable WRF + EWP science runs that could be performed at scale. We primarily tested these science runs in our Aristotle test environment where we deployed a single Docker container on a VM with 28 vCPUs, 240GB RAM, and Intel Xeon processors (Haswell or later).

It is worth noting that unlike AWS, the Red Cloud portion of Aristotle is not oversubscribed, so all mentions of vCPUs mean the user had full access to the underlying physical processor without competition. Since these runs did not require orchestration, deployment of the container the first time sufficed to provide a testbed for all future runs, eliminating deployment times for these jobs. An NFS share mount was bind-mounted into the Docker container for easier saving and back-up of results, and new jobs could be initiated via command-line or bash scripts.

*Table 4: WRF + EWP Container Single Node Scaling Runs*

| Runtime Results | | | Approx. Memory Usage | Domains | Simulation Data |
|---|---|---|---|---|---|
| real | user | sys | | | |
| 3m40.878s | 91m8.521s | 9m25.475s | * | 1 domain | 6 hours |
| 25m10.324s | 688m53.828s | 11m24.962s | 64GB | 2 domains | 6 hours |
| 137m58.096s | 3817m57.955s | 34m27.441s | 73.2GB | 3 domains | 6 hours |
| 256m27.356s | 7091m56.229s | 64m44.554s | 80GB | 4 domains | 6 hours |
| 385m56.337s | 10667m51.956s | 98m36.002s | 83GB | 5 domains | 6 hours |
| 8602m36.353s | 23916m31.170s | 1543m9.053s | * | 5 domains | 5.25 days |

*\* Note: Memory Usage was not captured for all runs*

The results in Table 4 provide a snapshot of some runs initiated by our atmospheric science collaborators in a single node scaling study performed to prepare for multi-node deployments of large jobs. The first five rows of the table detail results from varying the number of nested domains in the same simulation on 6 hours of weather data. The sixth row is a larger period of simulation data (5.25 days) with the same number of domains as the previous row. This last run is a good use case for the multi-node deployment scheme we have developed as part of this work.

**Numerical Stability**

The numerical stability of values computed by simulations run on HPC systems is an active area of research for various applications using numerical linear algebra in their computations. Due to the large number of interacting physics parameterization schemes active in WRF simulations, numerical stability of results from WRF are an important consideration. Results are often sensitive to the initial conditions chosen, grid size, boundaries, other system inputs, software and dependency versions, and underlying hardware of the system where simulations were run. Sometimes simulation results even vary when repeated on the same system with the same inputs.

Considering containers and the role they play in the numerical stability of these computations could likely yield some interesting and extensive studies of its own. For the purposes of this work, we barely scratched the surface of these possible analyses. We considered numerical stability in both the CONUS benchmark and science runs cases, both with merely a cursory look at graphs and statistics to ensure certain key variable values were within expected statistical ranges for reasonable variation.

For each new system and container run, we leveraged the Python scripts provided by NCAR with the CONUS benchmarks to compute the statistical differences between the expected results on NCAR systems and our system. Even considering that we may have used different compilers and dependency versions than NCAR when producing the benchmarks, our values were within acceptable limits of variance. One future step could be to do a more thorough analysis of these results across systems, and characterize what software or hardware changes, if any, account for the most variance when initial conditions and simulation sizes are kept standard.

For the scientific runs with our WRF + EWP Docker container, comparisons of numerical results of key variables were graphed by our collaborators for long-running simulations on Aristotle in Docker and bare-metal on another HPC system they have access to. Initially we believed that some variances were outside of acceptable ranges, however, a mistake in the initial conditions of a run were discovered through our analysis and rectified. Once re-run with the correct initial conditions, the results were within reasonable ranges of variance for different systems. Again, this is an opportunity for further study.

## 4.0 Conclusions

### 4.1 Discussion

Over the course of conducting this review of three container runtime technologies and Kubernetes orchestration, we detailed how adopting containers for the development and deployment of scientific computing applications raises both challenges and opportunities. We focused solely on the use of containers for scientific computing applications, with special attention to HPC applications, which present different complexities than what containers were originally developed for. The clearest takeaway we can draw from this study is that HPC application deployments using containers are typically much harder than simpler applications. They will likely require significant cyberinfrastructure development which can be time-consuming even for container-competent researchers. This burden on a research team, however, is significantly reduced when computational scientists and research software engineers work in concert with domain scientists to develop the appropriate solution for their use case.

**Challenges**

Before any code is built or executed, research teams who are thinking of using containers must first consider the overhead of initial container development. For domain scientists who are unaware of container technologies and best practices, or computational researchers who are new to the technology, learning and navigating the myriad choices of this technology's configurations as well as understanding their interactions and limitations can be daunting. The operational complexity when deploying containerized HPC-style codes at scale on supercomputers or Kubernetes clusters in the cloud requires deep knowledge to reliably execute. This fact, coupled with rapid changes to the technology, means significant time must be spent to become fully aware of this capability and keep up with it. This initial dismay can be somewhat mitigated by open source examples and community-provided support, but remains a challenge for many researchers.

There are a number of constraints that go into deciding which container runtime is the best option. We have investigated only three of the many possible container runtimes. These constraints include what is available in the computing environment that you are using, resource funding for various technologies and staff to develop the CI, technical support availability, existing group skills, and time available to learn the tools and technologies needed. Neither Docker Engine nor X-Containers was designed for scientific research whereas the Singularity container runtime is focused on scientific computing needs, which can ultimately make the decision for you. Flexibility to use the cloud significantly increases the appeal of Docker, especially if you are considering Kubernetes orchestration; however, whatever runtime one chooses, compromises abound.

We demonstrated the portability and consistency of containerized WRF simulation runs, a notoriously difficult application to compile and configure, on multiple platforms at various scales. We selected WRF as a particularly difficult application to containerize and deploy at scale because it is a good example of the challenges researchers face. Using containers for HPC application deployments that have specific implementations of MPI multi-process acceleration for multiple core and multiple node parallelization across hosts is much harder than using containers for simpler serial or single node applications. The WRF application remains a good candidate for containerization and automation to relieve researcher burdens despite the difficulties in configuring the dependent libraries, requested system resources, and other implementation factors, even on HPC systems with robust module ecosystems. The main challenge is the configuration of the MPI library and selecting the appropriate model for MPI communication at scale.

The target system itself can be a portability challenge as well. For HPC systems with specialized hardware such as Stampede2, portable container implementations are simply not available. This can be a significant deterrent in the initial development of a container; however, an application that is already containerized in a portable fashion may be less challenging to modify to conform to the target system if the process of building the container is repeatable.

There is a plethora of HPC applications with complicated build processes, dependency matching nightmares, and obtuse error messages that are frustrating to troubleshoot. Many could benefit from the reproducible environment provided by a container. The best person to develop a container is usually the person who is the most competent in installing software on different systems since they are well aware of target system intricacies. This person will likely need to collaborate closely with people who have container development skills. In general, more investment is needed in container development training if the benefits of containerization are to be realized.

During the prototyping phase of complex containerized software application development, the initial configurations can suffer delays requiring extensive troubleshooting, research, and sysadmin support. User support services at supercomputing centers have a great depth of experience in configuring and deploying complex scientific application codes, but their ability to securely support convenient container runtimes may be limited. This limitation may be mitigated somewhat by application-specific developer communities collaborating on public code infrastructure projects. It is our hope that our contribution of code examples and development of small-scale tests for feasibility and performance verification might lead to improved usability and advances in software tooling.

Once an HPC application has been developed in a portable fashion, and is even functional at scale, there are still application-specific concerns that must be addressed. Is the containerized deployment performant enough? Do the bottlenecks of the application prevent portability? Can numerical stability of results on different platforms be ensured for each containerized HPC application? These concerns must be addressed for each new container or community adoption may be thwarted.

**Opportunities**

Despite these challenges, once built, application containers often prove to be highly reusable, portable, and consistent execution environments that enable reproducible scientific workflows. This, in turn, reduces the computational support burden for both researchers and support staff. Once an application container is shared, its reusable, repeatable, and reproducible examples provided as open source (scripts, container build files, etc.) and public images can make an entire community of researchers more productive. Existing HPC application containers also assist in the development of new containers.

Once application containers are available to domain researchers who are trained or familiar with the workflow, they can spur faster deployment and refinement of workflows. For cases where cloud computing and other test environments are available (laptops, workstations, etc.) which support container runtimes, a portable container can be used by a developer in a variety of locations. Consequently, development workflows can skip the queue backlogs of busy HPC systems, giving scientists quicker access to results to iterate on. Furthermore, if developing in the cloud, these workflows can rapidly transition to larger jobs, reducing time to science.

While multi-node deployments can prove challenging for even competent computational researchers, it is worth noting that multi-node deployment is not always required in order to achieve very large application runs in the cloud. In some cases where cost is not prohibitive, the greater convenience of large VMs can make a single-node as productive as comparable multi-node cluster deployments. Though small by some standards, a single node is still able to provide more than 400 vCPUs [107] on some public cloud spot instance server types. They can also be migrated to dedicated on-demand Kubernetes clusters in the cloud and retain multi-node scale capabilities with a small percentage of runtime overhead.

## 4.2 Future Work

Beyond the existing opportunities presented by the current state of containers and orchestration, we see several possible avenues for future work to continue. Some of these ideas we are continuing to work on, and many of these could be larger studies performed by us or others in the community. The next major steps we see for this work involve continued development and use of our WRF containers, Kubernetes orchestration, and extension to other HPC applications.

**WRF Containers**

One improvement upon our existing WRF containers would be to produce a minimalistic version that is significantly scaled down in size, but is still reasonably portable and functional. In this study, we optimized for a functional development container that could be easily recompiled if needed. We have not found a need to recompile, and thus this functionality could be reduced. The way that a container is initially built for prototyping can be noticeably different from a trimmed down container than runs more efficiently at larger scales. However, the process we followed of scripting the builds within version control may be just as valuable for long term portability as container artifacts containing binaries themselves. Certainly, the public availability of this build process is more useful to the community than a binary that may or may not work on their system.

There are many ways in which it would be interesting and useful to the community to extend this study. The inclusion of more container runtimes, such as Shifter and Charliecloud, could broaden the scope of insights about HPC containers in particular. For our atmospheric science collaborators, Shifter would be of particular interest due to its availability on systems they have access to. A more detailed performance characterization and analysis of our WRF containers on these and other systems could provide a deeper understanding of the bottlenecks present across platforms. Further analysis of the numerical stability of WRF containers across platforms would also be valuable. And importantly, there are many more science use cases of WRF at scale that can be run and potentially leverage the CI we developed.

## Kubernetes Orchestration

Our Kubernetes orchestration techniques are ready to perform larger simulations at scale on AWS and GCP. Continued deployments of our Kubernetes clusters at scale could yield interesting results about the capabilities of cloud versus HPC clusters. A cost study of different Kubernetes on different cloud platforms and HPC resources would also be interesting.

Further developments of our orchestration technique still remain. For example, the Azure implementation of our Kubernetes cluster has yet to run an HPC application, which would be beneficial to inform further comparisons. We could also extend our existing Kubernetes infrastructure to try out other public cloud features such as using a Lustre filesystem on the cluster or testing various data storage configurations. Ultimately, trying alternative Terraform providers, and moving to a more stable Terraform configuration as it settles, will benefit our technique.

## Other HPC Containers and Orchestration

This is really just the beginning. We have only scratched the surface of the rich and complex space of HPC application containers and orchestration. Any of the work that we have done for WRF could be repeated for any number of other widely used HPC applications, and even less used but still important applications. We expect that many of these applications will experience some of the same containerization challenges that our application did, and it is our hope that the lessons learned and shared in this study will help the CI community overcome them.

## 5.0 Products

The Aristotle team and our collaborators implemented the following technologies and techniques and made them publicly available to the wider community:

- *Automated Deployment Methods* – implemented a Slurm HPC cluster in a cloud with OpenHPC 2 series based on CentOS/Rocky Linux 8 [108].
- *Radio Astronomy Container* – developed a single container of radio astronomy software [109] that combines the pipeline components developed for pulsar and other transient detections that can be deployed either on the cloud with Docker [110] or on an XSEDE HPC resource with Singularity [111].
- *Kubernetes Implementation Code for MPI Clusters* [112] – this code reflects the current state of experimental support for MPI applications managed by Terraform Kubernetes constructs that allow for automatic node count scaling and cloud portability. The software-based resource provisioning can best be attempted on AWS at the present time: it also works with Google Cloud Platform with MPI applications. Conversion to other cloud platforms should be possible with extensive changes to a platform-specific Terraform Kubernetes provider or other resource configurations. Familiarity with Kubernetes software concepts, resources provisioning on the desired cloud, and debugging parallel computing applications are recommended. The repository includes a "Getting Started with Kubernetes" tutorial as well.
- *WRF CONUS Benchmark Containers* [113] – implemented WRF 4.2.2 to run CONUS benchmarks on bare metal HPC in a Docker and a Singularity container.
- *WRF Docker Container* [114] – implemented a Docker container for WRF 3.8.1 with a Fitch patch.

A complete list of technologies and techniques [115] developed by the Aristotle Cloud Federation project and links to project publications [116] are available at the Aristotle portal [117].

## 6.0 References

[1]     Aristotle Cloud Federation (2021) Cornell University, University at Buffalo, and University of California, Santa Barbara [Online]. Available: https://federatedcloud.org/

[2]     Docker, Inc. (2021) Docker. Docker Inc. [Online]. Available: https://www.docker.com/

[3]     Aristotle Cloud Federation (2021) Use Case: Transient Detection in Radio Astronomy Data. Cornell University, University at Buffalo, and University of California, Santa Barbara [Online]. Available: https://federatedcloud.org/science/usecasetransients.php

[4]     Aristotle Cloud Federation (2021) Use Case: Application of Weather Research and Forecasting (WRF) Model for Climate-Relevant Simulations in the Cloud. Cornell University, University at Buffalo, and University of California, Santa Barbara [Online]. Available: https://federatedcloud.org/science/usecaseaerosols.php

[5]     G.M. Kurtzer, V. Sochat and M.W. Bauer, "Singularity: Scientific containers for mobility of compute," *PLOS ONE*, vol. 12, no. 5, pp. 1-20, 05 2017. [Online]. Available: https://journals.plos.org/plosone/article?id=10.1371/journal.pone.0177459

[6]     Z. Shen, Z. Sun, G-E Sela, E. Bagdasaryan, C. Delimitrou, R. van Renesse and H. Weatherspoon, "X-Containers: Breaking Down Barriers to Improve Performance and Isolation of Cloud-Native Containers," in *Proceedings of the 24th International Conference on Architectural Support for Programming Languages and Operating Systems*, ser. ASPLOS'19. New York, NY, USA: ACM, 2019, pp. 121-135 [Online]. Available: https://dl.acm.org/doi/10.1145/3297858.3304016

[7]     NCAR and UCAR (2021) WRF – The Weather Research and Forecasting Model. UCAR Mesoscale & Microscale Meteorology Laboratory [Online]. Available: https://www.mmm.ucar.edu/weather-research-and-forecasting-model

[8]     Docker Docs. (2021) Run the Docker daemon as a non-root user (Rootless mode). Docker Inc. [Online]. Available: https://docs.docker.com/engine/security/rootless/

[9]     Docker Hub (2021) Docker Inc. [Online]. Available: https://hub.docker.com/

[10]    Security in SingularityCE. (2021) Singularity Image Format (SIF). Skylabs Inc. & Project Contributors [Online]. Available: https://sylabs.io/guides/3.8/user-guide/security.html#singularity-image-format-sif

[11]    SchedMD (2021) Slurm Commercial Support and Development. SchedMD LLC [Online]. Available: https://www.schedmd.com/

[12]    Container Tools (2021) Singularity Hub Archive. GitHub [Online]. Available: https://singularityhub.github.io/

[13]    Sylabs (2021) Cloud Library. Sylabs Inc. [Online]. Available: https://cloud.sylabs.io/library

[14]    Sylabs. (2021) Support for Docker and OCI. Sylabs Inc. and Project Contributors [Online]. Available: https://sylabs.io/guides/3.8/user-guide/singularity_and_docker.html

[15]    Z. Shen, Z. Sun, G-E Sela, E. Bagdasaryan, C. Delimitrou, R. van Renesse and H. Weatherspoon, "X-Containers: Breaking Down Barriers to Improve Performance and Isolation of Cloud-Native Containers," in *Proceedings of the 24th International Conference on Architectural Support for Programming Languages and Operating Systems*, ser. ASPLOS'19. New York, NY, USA: ACM, 2019, pp. 121-135 [Online]. Available: https://dl.acm.org/doi/10.1145/3297858.3304016

[16]    Exotanium (2021) Exotanium, Inc. [Online]. Available: https://exotanium.io/

[17]    N. Zhou, Y. Gergiou, M. Pospieszny, L. Zhong, H. Zhou, C. Niethammer, B. Pejak, O. Marko and D. Hoppe, "Container orchestration on HPC systems through Kubernetes," *Journal of Cloud Computing*, 10, 16 (2021) [Online]. Available: https://journalofcloudcomputing.springeropen.com/articles/10.1186/s13677-021-00231-z

[18]    DevOpsSchool (2019) List of top container runtime interface projects, R. Kumar. DevOpsSchool, Inc. [Online]. Available: https://www.devopsschool.com/blog/list-of-top-container-runtime-interface-projects/

[19]    CapitalOne (2021) A Comprehensive Container Runtime Comparison. Capital One Financial Corporation [Online]. Available: https://www.capitalone.com/tech/cloud/container-runtime/

[20]    Kubernetes (2021) Container runtimes. The Kubernetes Authors [Online]. Available: https://kubernetes.io/docs/setup/production-environment/container-runtimes/

[21]    R. Knepper, S. Mehringer, A. Brazier, B. Barker and R. Reynolds, "Red Cloud and Aristotle: campus clouds and federations," in *Proceedings of Humans in the Loop: Enabling and Facilitating Research on Cloud Computing (HARC '19)*, New York, NY, USA, 4, pp. 1-6 [Online]. Available: https://dl.acm.org/doi/10.1145/3355738.3355755

[22]    Cornell University Center for Advanced Computing (2021) Cloud Computing Services – Red Cloud. Cornell University [Online]. Available: https://www.cac.cornell.edu/services/cloudservices.aspx

[23]    OpenStack (2021) Openstack.org [Online] Available: https://www.openstack.org/

[24]    Aristotle Cloud Federation (2021) Federation Resources. Cornell University, University at Buffalo, and University of California, Santa Barbara [Online]. Available: https://federatedcloud.org/using/federationresources.php

[25]    P. Vaillancourt, B. Wineholt, B. Barker, P. Deliyannis, J. Zheng, A. Suresh, A. Brazier, R. Knepper and R. Wolski, "Reproducible and Portable Workflows for Scientific Computing and HPC in the Cloud," in *Practice and Experience in Advanced Research Computing,* ser. PEARC '20. New York, NY, USA: ACM, 2020, pp. 311–320 [Online]. Available: https://doi.org/10.1145/3311790.3396659

[26]    P. Z. Vaillancourt, J. E. Coulter, R. Knepper and B. Barker, "Self-Scaling Clusters and Reproducible Containers to Enable Scientific Computing," *2020 IEEE High Performance Extreme Computing Conference (HPEC)*, pp. 1-8 [Online]. Available: https://ieeexplore.ieee.org/document/9286208

[27]    TOP500 The List (2021) TOP500 List – June 2021. TOP500.org [Online]. Available: https://www.top500.org/lists/top500/list/2021/06/

[28]    Stampede2 User Guide – TACC (2021) Using Modules to Manage Your Environment. Texas Advanced Computing Center, The University of Texas at Austin [Online]. Available: https://portal.tacc.utexas.edu/user-guides/stampede2#using-modules

[29]    AWS (2021) About AWS. Amazon Web Services, Inc. [Online]. Available: https://aws.amazon.com/about-aws/

[30]    Google Cloud (2021). Accelerate Your Transformation with Google Cloud. Google LLC. [Online]. Available: https://cloud.google.com/

[31]    Microsoft (2021) Azure. Microsoft Corporation [Online]. Available: https://azure.microsoft.com/en-us/

[32]    Kubernetes (2021) The Linux Foundation [Online]. Available: https://kubernetes.io/

[33]  AWS (2021) Amazon Elastic Kubernetes Service. Amazon Web Services, Inc. [Online]. Available: https://aws.amazon.com/eks/

[34]  Google Cloud (2021) Google Kubernetes Engine (GKE). Google LLC [Online]. Available: https://cloud.google.com/kubernetes-engine

[35]  Microsoft (2021) Azure Kubernetes Service (AKS). Microsoft Corporation [Online]. Available: https://azure.microsoft.com/en-us/services/kubernetes-service/

[36]  Kubernetes (2021 The Linux Foundation [Online]. Available: https://kubernetes.io/

[37]  DATADOG (2020) 11 Facts About Real-World Container Use. Datadog, Inc. [Online]. Available: https://www.datadoghq.com/container-report/

[38]  Linux NFS Overview, FAQ and HOW TO Documents (2021) VA Linux Systems sourceforge.net project [Online]. Available: http://nfs.sourceforge.net/

[39]  Terraform (2021) HashiCorp, Inc. [Online}. Available: https://www.terraform.io/

[40]  P. Vaillancourt, B. Wineholt, B. Barker, P. Deliyannis, J. Zheng, A. Suresh, A. Brazier, R. Knepper and R. Wolski, "Reproducible and Portable Workflows for Scientific Computing and HPC in the Cloud," in *Practice and Experience in Advanced Research Computing,* ser. PEARC '20. New York, NY, USA: ACM, 2020, pp. 311–320 [Online]. Available: https://doi.org/10.1145/3311790.3396659

[41]  GitHub (2021) federatedcloud/kubernetes-mpi-clusters. GitHub, Inc. [Online]. Available: https://github.com/federatedcloud/kubernetes-mpi-clusters

[42]  Docker Docs (2021) What can we help you with. Docker, Inc. [Online}. https://docs.docker.com/

[43]  Docker Docs. (2021) Best practice for writing Dockerfiles. Docker Inc. [Online]. Available: https://docs.docker.com/develop/develop-images/dockerfile_best-practices/

[44]  Containers@TACC (2021) Introduction to Singularity. Texas Advanced Computing Center, The University of Texas at Austin [Online]. Available: https://containers-at-tacc.readthedocs.io/en/latest/singularity/01.singularity_basics.html

[45]  Sylabs (2021) Support for Docker and OCI: Overview. Sylabs Inc. and Project Contributors [Online]. Available: https://sylabs.io/guides/3.7/user-guide/singularity_and_docker.html

[46]  Sylabs (2021) Troubleshooting. Sylabs Inc. and Project Contributors [Online]. Available: https://sylabs.io/guides/3.7/user-guide/singularity_and_docker.html#troubleshooting

[47]  Docker Hub (2021) cornell/cac/wrf. Docker, Inc. [Online]. Available: https://hub.docker.com/r/cornellcac/wrf/tags?page=1&ordering=last_updated

[48]  Sylabs (2021) Troubleshooting. Sylabs Inc. and Project Contributors [Online]. Available: https://sylabs.io/guides/3.7/user-guide/singularity_and_docker.html#troubleshooting

[49]  GitHub (2021) singularityhub/docker2singularity. GitHub, Inc. [Online]. Available: https://github.com/singularityhub/docker2singularity

[50]  Sylabs (2021) Build Image from Scratch. Sylabs Inc. & Project Contributors [Online]. Available: https://sylabs.io/guides/3.4/user-guide/quick_start.html#build-images-from-scratch

[51]  Z. Shen, Z. Sun, G-E Sela, E. Bagdasaryan, C. Delimitrou, R. van Renesse and H. Weatherspoon, "X-Containers: Breaking Down Barriers to Improve Performance and Isolation of Cloud-Native Containers," in *Proceedings of the 24th International Conference on Architectural Support for Programming Languages and Operating*

*Systems*, ser. ASPLOS'19. New York, NY, USA: ACM, 2019, pp. 121-135 [Online]. Available: https://dl.acm.org/doi/10.1145/3297858.3304016

[52]     Docker Docs. (2021) Use the Device Mapper storage driver. Docker Inc. [Online]. Available: https://docs.docker.com/storage/storagedriver/device-mapper-driver/

[53]     Google Cloud (2021) Compute Engine Guides: Creating, deleting, and deprecating custom images. Google LLC [Online]. Available: https://cloud.google.com/compute/docs/images/create-delete-deprecate-private-images

[54]     GitHub (2021) hasicorp/terraform-provider-kubernetes. GitHub, Inc. [Online}. Available: https://github.com/hashicorp/terraform-provider-kubernetes

[55]     AWS (2021) What is Amazon VPC? Amazon. Amazon Web Services, Inc. [Online]. Available: https://docs.aws.amazon.com/vpc/latest/userguide/what-is-amazon-vpc.html

[56]     Kubeflow (2021) MPI Training (MPIJob). The Kubeflow Authors [Online]. Available: https://www.kubeflow.org/docs/components/training/mpi/

[57]     TOP500 (2021) The Linpack Benchmark. TOP500.org [Online]. Available: https://www.top500.org/project/linpack/

[58]     GitHub (2021) Aristotle Cloud Federation, Cornell University. GitHub, Inc. [Online]. Available: https://github.com/federatedcloud

[59]     R. White and H. Christensen (2017). ROS and Docker. In: A. Koubaa (eds) Robot Operating System (ROS), in Computational Intelligence, vol 707, Springer, Cham. [Online]. Available: https://link.springer.com/chapter/10.1007/978-3-319-54927-9_9

[60]     Docker (2021) Docker Registry. Docker Inc. [Online]. Available: https://docs.docker.com/registry/

[61]     AWS (2021) AWS Graviton Processor. Amazon Web Services, Inc. [Online]. Available: https://aws.amazon.com/ec2/graviton/

[62]     Wikipedia (2021) X86-64. Wikimedia Foundation, Inc. [Online].  Available: https://en.wikipedia.org/wiki/X86-64

[63]     GitHub (2021) sylabs/singularity-cri. GitHub, Inc. [Online]. Available: https://github.com/sylabs/singularity-cri

[64]     Lustre (2021) Welcome to the official home of the Lustre filesystem. OpenSFS and EOFS [Online]. Available: https://www.lustre.org/

[65]     C. Arango, R. Dernat, and J. Sanabria, "Performance Evaluation of Container-based Virtualization for High Performance Computing Environments," in *arXiv*, 2018 [Online]. Available: https://arxiv.org/abs/1709.10140

[66]     P. Saha, A. Beltre, P. Uminski and M. Govindaraju, "Evaluation of Docker Containers for Scientific Workloads in the Cloud," in *arXiv,* 2019 [Online]. Available: https://arxiv.org/abs/1905.08415

[67]     A.J. Young, K. Pedretti, R.E. Grant and R. Brightwell, "A Tale of Two Systems: Using containers to Deploy HPC Applications on Supercomputers and Clouds," *2017 IEEE International Conference on Cloud Computing Technology and Science (CloudCom)*, 2017, pp. 74-81 [Online]. Available: https://ieeexplore.ieee.org/document/8241093

[68]     C. Arango, Remy Dermet and J. Snabria, "Performance Evaluation of Container-based Virtual Environments," in *arXiv*, (2017) [Online]. Available: https://arxiv.org/abs/1709.10140

[69]    A.M. Joy, "Performance comparison between Linux containers and virtual machines," 2015 International Conference on Advanced in Computer Engineering and Applications, 2015, pp. 342-346 [Online]. Available: https://ieeexplore.ieee.org/document/7164727

[70]    P. Saha, A. Beltre, P. Uminski and M. Govindaraju, "Evaluation of Docker Containers for Scientific Workloads in the Cloud," in *Proceedings of the Practice and Experience on Advanced Research Computing (PEARC '18),* 2018, pp. 1-8 [Online]. Available: https://dl.acm.org/doi/10.1145/3219104.3229280

[71]    K. Liu, K. Aida, S. Yokoyama and Y. Masatani, "Flexible Container-Based Computing Platform on Cloud for Scientific Workflows, *2016 International Conference on Cloud Computing Research and Innovations (ICCCRI)*, 2016, pp. 56-63 [Online]. Available: https://ieeexplore.ieee.org/document/7600178

[72]    R. Morabito, J. Kjallman and M. Komu, "Hypervisors vs. Lightweight Virtualization: A Performance Comparison," *2015 IEEE International Conference on Cloud Engineering*, 2015, pp. 386-393 [Online]. Available: https://ieeexplore.ieee.org/document/7092949

[73]    R.R. Yadav, E.T.G. Sousa and G.R.A. Callou, "Performance Comparison Between Virtual Machines and Docker Containers," in *IEEE Latin America Transactions*, vol 16, no. 8, pp. 2282-2288, Aug. 2018 [Online}. Available: https://ieeexplore.ieee.org/document/8528247

[74]    NCAR and UCAR (2021) WRF – The Weather Research and Forecasting Model. UCAR Mesoscale & Microscale Meteorology Laboratory [Online]. Available: https://www.mmm.ucar.edu/weather-research-and-forecasting-model

[75]    GitHub (2021) TACC/benchtool. GitHub, inc. [Online]. Available: https://github.com/TACC/benchtool

[76]    NCAR and UCAR (2021) WRF – The Weather Research and Forecasting Model. UCAR Mesoscale & Microscale Meteorology Laboratory [Online]. Available: https://www.mmm.ucar.edu/weather-research-and-forecasting-model

[77]    Cornell University, Earth and Atmospheric Sciences (2021) Sara C. Pryor [Online]. Available: https://www.eas.cornell.edu/faculty-directory/sara-c-pryor

[78]    Aristotle Cloud Federation (2021) Use Case: Application of Weather Research and Forecasting (WRF) Model for Climate-Relevant Simulations in the Cloud. Cornell University, University at Buffalo, and University of California, Santa Barbara [Online]. Available: https://federatedcloud.org/science/usecaseaerosols.php

[79]    Cornell Chronicle (2020) Grant supports development of efficient offshore wind farms. Cornell University [Online]. Available: https://news.cornell.edu/stories/2020/10/grant-supports-development-efficient-offshore-wind-farms

[80]    P.J.H. Volker, J. Badfer, A.N. Hahmann, and S. Ott, "The Explicit Wake Parametrisation V.1.0: a wind farm parametrization in the mesoscale model WRF," Geoscientific Model Development 8, pp. 3481-3522 [Online]. Available: https://pdfs.semanticscholar.org/d5a7/10396d9465059d4d0a4cea28b92d285557a1.pdf

[81]    WRF Users Page (2021) UCAR Mesoscale & Microscale Meteorology Laboratory [Online]. Available: https://www2.mmm.ucar.edu/wrf/users/index.html

[82]    WRF v.4.2.2 Benchmark Cases: CONUS 12-km and CONUS 2.5-km (2021) UCAR Mesoscale & Microscale Meteorology Laboratory [Online]. Available: https://www2.mmm.ucar.edu/wrf/users/benchmark/benchdata_v422.html

[83]   GitHub (2021) NCAR/WRF DOCKER. GitHub, Inc. [Online]. Available: https://github.com/NCAR/WRF_DOCKER

[84]   How to Compile WRF: The Complete Process (2021) UCAR Mesoscale & Microscale Meteorology Laboratory [Online]. Available: https://www2.mmm.ucar.edu/wrf/OnLineTutorial/compilation_tutorial.php

[85]   WRF & MPAS-A Support Forum (2021) UCAR Mesoscale & Microscale Meteorology Laboratory [Online]. Available: https://forum.mmm.ucar.edu/phpBB3/

[86]   GitHub (2021) federatedcloud/Docker-WRF-3.8.10-Fitch. GitHub, Inc. [Online]. Available: https://github.com/federatedcloud/Docker-WRF-3.8.1-Fitch

[87]   GitHub (2021) federatedcloud/WRFv4-Benchmarking. GitHub, Inc. [Online]. Available: https://github.com/federatedcloud/WRFv4-Benchmarking/tree/main/Stampede2

[88]   GitHub (2021) federatedcloud/WRFv4-Benchamrking. GitHub, Inc. [Online]. Available: https://github.com/federatedcloud/WRFv4-Benchmarking

[89]   Intel (2021) Intel oneAPI HPC Toolkit. Intel Corporation [Online]. Available: https://software.intel.com/content/www/us/en/develop/tools/oneapi/hpc-toolkit.html#gs.aql0n4

[90]   Intel (2021) Intel oneAPI HPC Toolkit Container [Online]. Available: https://software.intel.com/content/www/us/en/develop/articles/containers/oneapi-hpc-toolkit.html

[91]   GitHub (2021) WRFv4-Benchmarking/Docker. GitHub, Inc. [Online]. Available: https://github.com/federatedcloud/WRFv4-Benchmarking/tree/main/Docker

[92]   Docker Hub (2021) cornellcac/wrf. Docker, Inc. [Online]. Available: https://hub.docker.com/u/cornellcac

[93]   Docker Hub (2021) cornellcac/wrf TAG 4.2.2-intel-7415915e0b8e. Docker, Inc. [Online]. Available: https://hub.docker.com/r/cornellcac/wrf/tags?page=1&ordering=last_updated

[94]   Containers@TACC (2021) Message Passing Interface (MPI) for running on multiple nodes. Texas Advanced Computing Center, The University of Texas at Austin [Online]. Available:https://containers-at-tacc.readthedocs.io/en/latest/singularity/03.mpi_and_gpus.html#message-passing-interface-mpi-for-running-on-multiple-nodes

[95]   Sylabs (2021) Singularity and MPI Applications. Sylabs Inc. and Project Contributors [Online]. Available: https://sylabs.io/guides/3.7/user-guide/mpi.html

[96]   GitHub (2021) TACC/tacc-containers. GitHub, Inc. [Online]. Available: https://github.com/TACC/tacc-containers

[97]   Containers@TACC (2021) Base Docker images. Texas Advanced Computing Center, The University at Texas at Austin [Online]. Available: https://containers-at-tacc.readthedocs.io/en/latest/singularity/03.mpi_and_gpus.html#base-docker-images

[98]   GitHub (2021) federatedcloud/kubernetes-mpi-clusters. GitHub, Inc. [Online]. Available: https://github.com/federatedcloud/kubernetes-mpi-clusters

[99]   GitHub (2021) federatedcloud/kubernetes-mpi-clusters. GitHub, Inc. [Online]. Available: https://github.com/federatedcloud/kubernetes-mpi-clusters

[100]   Terraform Registry (2021) Getting Started with AWS EKS. HashiCorp, Inc. [Online]. Available: https://registry.terraform.io/providers/hashicorp/aws/2.34.0/docs/guides/eks-getting-started

[101]   Intel (2021) Intel VTune Profiler User Guide. Intel Corporation [Online]. Available: https://software.intel.com/content/www/us/en/develop/documentation/vtune-help/top.html

[102]   Intel (2021) Intel VTune Profiler User Guide: MPI Code Analysis. Intel Corporation [Online]. Available: https://software.intel.com/content/www/us/en/develop/documentation/vtune-help/top/analyze-performance/code-profiling-scenarios/mpi-code-analysis.html

[103]   Intel (2021) Intel VTune Profiler Analysis: Cookbook. Intel Corporation [Online]. Available: https://software.intel.com/content/www/us/en/develop/documentation/vtune-cookbook/top/configuration-recipes/profiling-mpi-applications.html

[104]   TACC User Portal (2021) Remora – Resource Monitoring for Remote Applications. Texas Advanced Computing Center, The University of Texas at Austin [Online]. Available: https://portal.tacc.utexas.edu/software/remora

[105]   AWS (2021) Amazon EC2 C5 Instance. Amazon Web Services, Inc. [Online]. Available: https://aws.amazon.com/ec2/instance-types/c5/

[106]   AWS (2021) Amazon EC2 M5 Instance. Amazon Web Services, Inc. [Online]. Available: https://aws.amazon.com/ec2/instance-types/m5/

[107]   Google Cloud (2021) Machine families. Google, LLC. [Online]. Available: https://cloud.google.com/compute/docs/machine-types

[108]   GitHub (2021) federatedcloud/wtf-cluster-openstack. GitHub, Inc. [Online]. Available: https://github.com/federatedcloud/wrf-cluster-openstack

[109]   GitHub (2021) federatedcloud/pulsar-pipeline-container. GitHub, Inc. [Online]. Available: https://github.com/federatedcloud/pulsar-pipeline-container

[110]   Docker Hub (2021) cornellcac/pulsar-pipeline. Docker, Inc. [Online]. Available: https://hub.docker.com/r/cornellcac/pulsar-pipeline

[111]   DataLad (2021) datasets.datalad.org/shub/federatedcloud/pulsar-pipeline-container. DataLad Team [Online]. Available: https://datasets.datalad.org/?dir=/shub/federatedcloud/pulsar-pipeline-container

[112]   GitHub (2021) federatedcloud/kubernetes-mpi-clusters. GitHub, Inc. [Online]. Available: https://github.com/federatedcloud/kubernetes-mpi-clusters

[113]   GitHub (2021) federatedcloud/WRFv4-Benchamrking. GitHub, Inc. [Online]. Available: https://github.com/federatedcloud/WRFv4-Benchmarking

[114]   GitHub (2021) federatedcloud/Docker-WRF-3.8.1-Fitch. GitHub, Inc. [Online]. Available: https://github.com/federatedcloud/Docker-WRF-3.8.1-Fitch

[115]   Aristotle Cloud Federation (2021) Technologies. Cornell University, University at Buffalo, and University of California, Santa Barbara [Online]. Available: https://federatedcloud.org/using/technologies.php

[116]   Aristotle Cloud Federation (2021) Publications. Cornell University, University at Buffalo, and University of California, Santa Barbara [Online]. Available: https://federatedcloud.org/about/publications.php

[117]   Aristotle Cloud Federation (2021) Cornell University, University at Buffalo, and University of California, Santa Barbara [Online]. Available: https://federatedcloud.org/